\documentclass[
preprint,
aps,prd,
nofootinbib,
superscriptaddress,
showpacs,
tightenlines
]{revtex4}
\usepackage{amsmath}
\usepackage{amssymb}
\usepackage{bm}
\usepackage[dvips]{color,graphicx}
\begin{document}

\title{Generalized Parton Distributions from Hadronic Observables: Zero Skewness}

\author{Saeed Ahmad}
\email[E-mail: ]{sa8y@virginia.edu}
%\affiliation{University of Virginia, 382 McCormick Road, Charlottesville, Virginia 22904, USA.}
%
\author{Heli Honkanen} 
\email[E-mail: ]{hh9e@virginia.edu}
%\affiliation{University of Virginia, 382 McCormick Road, Charlottesville, Virginia 22904, USA.}
%
\author{Simonetta Liuti} 
\email[E-mail: ]{sl4y@virginia.edu}
%\affiliation{University of Virginia, 382 McCormick Road, Charlottesville, Virginia 22904, USA.}
%
\author{Swadhin~K.~Taneja} 
\email[E-mail: ]{taneja@cpht.polytechnique.fr}

\altaffiliation[Present address: ]{Ecole Polytechnique, CPHT, F91128 Palaiseau Cedex, France}
\affiliation{University of Virginia, 382 McCormick Road, Charlottesville, Virginia 22904, USA.}


\begin{abstract}
We propose a physically motivated parametrization for the unpolarized 
generalized parton distributions.
At zero value of the skewness variable, $\zeta$, the parametrization is 
constrained by simultaneously fitting the experimental data on both 
the nucleon elastic form factors and the deep inelastic structure functions. 
A rich phenomenology can be addressed based on this parametrization. 
In particular, we track the behavior of the 
average: {\it i)} interparton distances as a function of the momentum 
fraction, $X$, 
{\it ii)} $X$ as a function of the four-momentum transfer, $t$; 
{\it iii)} the intrinsic 
transverse momentum $k_\perp$ as a function of $X$.      
We discuss the extension of our parametrization to $\zeta \neq 0$ where 
additional constraints are provided by higher moments of the generalized
parton distributions obtained from {\it ab initio} lattice QCD calculations.
\end{abstract}

\maketitle

\baselineskip 3.5ex
%%%%%%%%%%%%%%%%%%
\section{Introduction}
Most information on the quark and gluon 
structure of hadrons has come so far 
from inclusive Deep Inelastic 
Scattering (DIS) type experiments. With an appropriate selection of 
probes and reactions, 
accurate measurements conducted through the years  
allowed one to map out in detail the different components of proton 
structure, the Parton Distribution Functions (PDFs) 
in a wide kinematical region of  
the four-momentum transfer, $Q^2$, and of the longitudinal momentum 
fraction of the proton's momentum, $x_{Bj}=Q^2/2M\nu$, $\nu$ being the energy transfer and 
$M$ the proton mass. 

Recently, a whole new dimension was added to our understanding of hadronic structure, with
the observation that in a number of exclusive experiments one can, on one side, study a 
wider range of flavor and spin dependent 
combinations of PDFs with respect to those obtained from inclusive scattering, and 
on the other, new, qualitatively different information can in principle be extracted 
from a specific class of experiments, including Deeply Virtual Compton
Scattering (DVCS), and hard Exclusive Meson Production (EMP).
The information from these processes is coded in terms of  ``off-forward'' contributions, 
or the Generalized Parton Distributions (GPDs) \cite{DMul1,Ji1,Rad1}.
GPDs allow us to access partonic configurations 
with a given longitudinal momentum fraction, similarly to DIS, but 
also at a specific (transverse) location inside the hadron \cite{Bur}. 

PDFs are extracted directly from inclusive measurements 
of the DIS structure functions
at a given $Q^2$.  Perturbative QCD (PQCD) evolution   
connects the PDF values at $Q^2$ with the ones at an initial scale, $Q_o^2$.     
The initial PDFs
are usually given in parametric forms that broadly reproduce  
the behavior expected in a few limiting cases: 
they include , for instance, a Regge-type behavior 
at low $x_{Bj}$, and a quark-counting type behavior in the limit $x_{Bj} \rightarrow 1$.
A number of sum rules such as Adler's, and the momentum sum
rule provide additional constraints.
It is also generally understood that the proton is an ``emptier'' object dominated
by its minimal -- valence -- components at low 
$Q^2$. 
%Q
The sea quark content is determined both by quark-antiquark pairs and gluons radiations 
which characterize perturbative evolution from the initial low scale,  
and by ``intrinsic''  components also expected to be present at low scales.
PDF parametrizations
have become more sophisticated through the years both because of the 
continuous addition of DIS data in increasingly extended kinematical regimes, 
and because of phenomenological developments allowing one to extend 
the number of hard processes from which PDFs can be extracted. 
It is now possible to describe the proton structure functions with relative accuracy 
 in the regime: 
$10^{-4} \lesssim x_{Bj} \lesssim 0.75$, and $1 \lesssim Q^2 \lesssim 10^4$ GeV$^2$ \cite{PDG}. 

The matching between measured quantities and leading order predictions   
for DVCS/EMP and GPDs should proceed, in principle, 
similarly to the inclusive case, in view of the 
factorization theorem discussed in Ref.~\cite{Ji1}. 
There are however a few important caveats due to the fact that 
GPDs describe the non-perturbative contribution to an {\em amplitude}.
%where the struck quark
%after emitting a real photon (or a meson) is reabsorbed by the proton with a
%different momentum fraction.
Both the real and imaginary parts of the amplitude are physical observables in 
the $\gamma^* P \rightarrow \gamma P^\prime$ process, obtained from
the interference term for the DVCS and 
Bethe-Heitler (BH) processes (see \cite{BelDVCS} and reviews in \cite{Die_rev,BelRad}). 

The leading order amplitude for DVCS is shown in Fig.~\ref{fig1} along with the 
relevant kinematical variables, namely: the longitudinal momentum fraction 
taken by the initial quark,
$X$, $Q^2 \equiv -q_\mu^2$, the four-momentum transfer squared between the initial 
and final proton states, $t \equiv -\Delta^2$, and the longitudinal momentum
transfer fraction of the initial proton momentum, the so-called ``skewness'', $\zeta$. 
Because of the extra two parameters, $\zeta$ and $t$, obtaining initial parametrizations 
in a similar fashion as for inclusive parton 
distributions is a formidable task.     
Furthermore, experimental measurements of GPDs are remarkably 
more complicated than in inclusive 
DIS, essentially due to the exclusive nature of DVCS/EMP.
% and to the  coherent superposition with the Bethe-Heitler process.
Despite the high  performance of current facilities such as 
Jefferson Laboratory, one cannot realistically expect in the near future 
a similar amount and quality of data as in 
inclusive experiments. 
%Despite the number of experiments performed and planned for 
%the near future, it will be therefore a long 
%time before one can reach both the accuracy and kinematical
%coverage of DIS measurements.

Given the importance of the physics underlying GPDs, and the far reaching consequences
of their study   
related to a number of key questions such as the contributions of the partons 
angular momentum to the nucleon's spin, the exploration of 3D spatial 
images of the nucleon, and the connection to Transverse Momentum Distributions (TMDs) 
\cite{Die_rev,BelRad}, 
it is most important at the present stage, to explore 
whether currently available inclusive  
data can also provide additional constraints on GPDs. These constraints can 
both supplement the, so far, scarce experimental
results obtained directly from DVCS/EMP, and at the same time provide a 
guidance for future precision exclusive experiments
at both Jefferson Lab at 12 GeV, and future colliders. 

Experimental constraints from other-than-DVCS-type data on GPDs are obtained from: 
{\it i)}  The nucleon form factors providing 
integrals of GPDs over $X$ at a fixed $t$; 
{\it ii)} The PDFs, representing 
the $(\zeta, t \rightarrow 0)$ limit of GPDs. 
An additional check is also provided by the relation, through simple Fourier 
transformation, between zero skewedness
GPDs and Impact Parameter Dependent PDFs (IPPDFs) \cite{Bur}.
Physically meaningful GPDs should in fact reproduce the correct behavior 
of partonic configuration radii obtained from their impact parameter space representation.

Because both the form factors and PDFs are independent of $\zeta$, these constraints apply 
exclusively to 
the case $\zeta =0$, $t \equiv -\Delta_\perp^2$.
Zero skewness GPDs extracted using available experimental constraints 
were first considered in the initial 
phenomenological studies of Refs.~\cite{Sto,GSL,afa,LiuTan1}.
Fully quantitative fits were subsequently performed in Refs.~\cite{DieKro1,VandH1}. 
%%%
Defining a parametrization at $\zeta \neq 0$, however, requires
the additional condition of polynomiality to be satisfied \cite{Die_rev}
Moreover, a physical interpretation in terms of partonic components 
becomes less transparent both at $X=\zeta$ -- the  
``stopped returning quark'' region -- and at  $X<\zeta$ where the dominating process
is scattering from a $q \overline{q}$ pair emerging from the initial nucleon. 
The only guidance for a parametrization at $\zeta \neq 0$   
has been provided, so far, by the Double Distribution (DD) hypothesis \cite{RadDD,VGG},
that has a built-in property of polynomiality. More recently, 
the Mellin-Barnes integral representation \cite{MuelScha}, and 
the dual representation \cite{GuzPol} were proposed, where the GPDs 
were obtained within a generalization of the anti-Mellin transform 
approach used for PDFs.  
Nevertheless, similarly to what found for DIS \cite{Fox,Yndurain}, 
the extraction of GPDs from moments 
formally requires a continuation to complex $n$ 
that, because of an oscillating term inherent in the moments integral, 
can be a source of ambiguities.
This has so far hampered an accurate extraction using other than 
simplified models.

Motivated by this situation, in this paper, we introduce a practical method
to extract GPDs from experimental data, that can be extended also at $\zeta \neq 0$, 
by using  additional $\zeta$-dependent constraints from {\it ab-initio} lattice 
QCD calculations of the first three moments of GPDs \cite{zan,LHPC}.
% in addition to the $\zeta=0$ constarints {\it i), ii)}, described above.
We stress that differently from the DDs, representing a model calculation, 
our approach is, for the first time to our knowledge, 
an attempt to obtain a realistic parametrization. 
%making use of experimental data in combination with lattice results. 
Given the paucity of current direct experimental measurements of GPDs, 
our goal is to provide more stringent, model independent  
predictions that will be useful both for model builders, in order 
to understand the dynamics of GPDs, and for the planning of future 
DVCS type experiments. 

Our paper organization is as follows: In Section \ref{sec2} we describe our approach for 
a physically motivated parametrization valid both in the $\zeta=0$ and $\zeta \neq 0$ cases. 
We consider the unpolarized GPDs, $H$ and $E$, 
for which the lattice moments are the most accurate.
In Sections \ref{sec3}, and \ref{sec4} we discuss in detail the $\zeta =0$ case: in III we perform a detailed 
comparison with the data on both form factors and PDFs; 
In IV we show the 
phenomenology of GPDs at $\zeta=0$ by illustrating the role of the various 
quantities: the quarks transverse radii $\langle y_q \rangle$, the intrinsic transverse momentum 
${\bf k}_\perp$, the average $X$ values contributing to the nucleon form factors 
as a function of $t$. We also discuss the feasibility of the extraction using lattice results 
and introduce our method.
For ease of presentation, a number of graphs and more quantitative results on the $\zeta \neq 0$ 
case are reported in a following paper \cite{part2}. 
In Section V we draw our conclusions.

%%%%%%%%
%%%%%%%% SECTION II
%%%%%%%%
\section{A Physically Motivated Parametrization for Unpolarized GPDs} 
\label{sec2}
%We start by describing the connection between GPDs and both the measured 
%observables -- the nucleon form factors --
%and the lattice moments used to constrain our parametrization in the enitre
%range of the skewedness parameter.
 
GPDs parameterize the non-perturbative vertex in the DVCS process depicted in Fig.~\ref{fig1}.   
Scattering from an unpolarized proton (neutron) is described by two
independent GPDs: $H$, and $E$, from the vector ($\gamma_\mu$)
and tensor ($\sigma_{\mu\nu}$) interactions, respectively, 
that depend on three kinematical invariants, besides
the initial photon's virtuality, $Q^2$: the longitudinal momentum transfer,
$\zeta = Q^2/2(Pq)$, the four-momentum transfer squared, $\Delta^2=-t$, and the
variable $X=(kq)/(Pq)$, representing the Light Cone (LC) momentum fraction carried by the struck 
parton with momentum $k$. 
The relations between the variables used in this paper and the analogous set of kinematical variables 
in the  ``symmetric'' system, frequently used in the literature are given along with 
the definitions of the hadronic tensors components in Refs.~\cite{MarGol}.  
Note the slightly different choice from Ref.~\cite{BroDie} 
for which the sea quarks GPDs are also defined at $X>0$
(see Appendix A). 

At present, due to both the small experimental coverage mentioned above, and to
the somewhat less transparent physical interpretation of the observables, 
an important role in the 
shaping of a parametrization for GPDs is played by the intuition on 
the underlying non-perturbative dynamics. 

In the general case of non-zero skewedness, one distinguishes
two kinematical regions: {\it i)} $\zeta<X<1$, where
the dominant process is where a quark from
the initial proton with LC momentum fraction $X$,
is struck by the initial photon and is subsequently reabsorbed in the proton (with $X-\zeta>0$);
{\it ii)} $0<X<\zeta$, where the final quark propagates backwards, or, 
correspondingly, a quark-antiquark pair from the
initial proton, with an asymmetric partition of LC momenta, 
participates in the scattering process.
%also called the DGLAP region because it is amenable
%to ``standard'' perturbative QCD evolution, %as it will be clarified later, 

The physics of the two regions is most easily understood within 
a field theoretical description where the lowest order is given 
by a covariant quark-nucleon scattering 
amplitude, with the nucleon-quark-diquark vertex being 
a Dirac matrix multiplied by a scalar function.
%%%
\footnote{The most general case involves 
a linear combination of Dirac matrices \protect\cite{MeyMul,MelSchTho})}. 
Physical intuition on DVCS processes can then be obtained 
by viewing the covariant diagrams as the sum of all possible 
time ordered diagrams.   
%The contribution of each time ordered diagram is inversely 
%proportional to the energy difference: 
%\[ E_i - E_{int}, \]
%\noindent 
%$E_i$ and $E_{int}$ being the sum of the energies of all particles in the initial  and intermediate
%states respectively. 
The invariant amplitude for DVCS corresponds to $4$!  time ordered diagrams. 
They are grouped into the following classes of processes (Fig.~\ref{fig2}):
{\bf (a)} all particles moving forward (Fig.~\ref{fig2}a); 
{\bf (b)} the initial photon splits into a quark-antiquark pair that then interacts with the hadronic
system; {\bf (c)} an initial quark-antiquark pair originates from the initial proton and scatters 
from the probe (Fig.~\ref{fig2}b). Each process has a corresponding ``crossed term'' (not drawn in the
figure). 

We consider the following choice of frame, and four-momentum components:%
%%%%
\footnote{We use the notation $a^\mu \equiv (a^0 \equiv E_a; {\bf a}_\perp, a^3)$, and
$a^{\pm} = (a^0 \pm a^3)/\sqrt{2}$.} 
%%%%
\begin{subequations}
\begin{eqnarray}
\label{kinematics}
q & \equiv & (0; {\bf q}, 0)  \\
%%%%%% 
P & \equiv & (P + \frac{M^2}{2P}; {\bf 0}, P) \\
%%%%%%% 
k & \equiv & \left(X P + \frac{{\bf k}_\perp^2 + m_q^2}{2XP}; {\bf k}_\perp, XP \right)\\ 
%%%%%
\label{kx}
k_X & \equiv & \left((1-X) P + \frac{{\bf k}_\perp^2 + M_X^2}{2(1-X)P}; -{\bf k}_\perp, (1-X)P \right)\\
%%%%%%% 
k^\prime & \equiv & \left((X-\zeta) P + 
\frac{({\bf k}_\perp - {\bf \Delta}_\perp)^2 + m_q^2}{2(X-\zeta)P}; {\bf k}_\perp - \Delta_\perp, (X -\zeta) P \right)\\ 
%%%%%%%%%%%%
P^\prime & \equiv & \left( (1-\zeta) P + 
\frac{{\bf \Delta}_\perp^2 + M^2}{2(1-\zeta)P}; -{\bf \Delta}_\perp, (1 -\zeta) P \right)  \\ 
%%%%%%%%%%
\Delta  & \equiv & \left( \zeta P + \frac{-t+{\bf \Delta}_\perp^2}{2 \zeta P}; {\bf \Delta}_\perp, \zeta P \right)
\end{eqnarray}
\end{subequations}
where, moreover, $q^\prime  =  q + \Delta$.
Eq.~(\ref{kx}) gives the components of 
a spectator system, {\it i.e.} a {\it diquark},
that is our main assumption in the 
following sections, namely that the spectral distribution of states 
appearing in principle in the quark correlator can be replaced by one
state with a given mass.  
Finally, in the given frame, contributions
dominated by the hadronic components of the initial photon (case {\bf (c)}) 
are absent. More generally, all processes where one of the particles is moving
backwards (or one of the particles has a longitudinal momentum opposite to the protons' one)
vanish as inverse powers of $P$. 

%%%%%%%%
\subsection{The  $X>\zeta$ region}
\label{sec2a}
%%%%%%%%
At $X>\zeta$,    
the proton splits into a quark carrying a LC momentum fraction $X=k^+/P^+$, transverse momentum
${\bf k}_\perp$, and a spectator
system with $1-X,  = k_X^+/P^+$, and $-{\bf k}_\perp$ (Fig.~\ref{fig2}a). 
Similarly the right side vertex describes the coalescence of the final quark and 
the spectator system into an outgoing proton (all particles are moving forward). 
With an appropriate choice for the 
$P \rightarrow (k \, k_X)$ and $P^\prime \rightarrow (k^\prime \, k_X)$ vertices, {\it i.e.} assuming a 
spectator diquark with both scalar and axial vector
components, the DVCS matrix element, $F(X,\zeta,t)$, can be written at leading
order in $Q^2$, as (see also Ref.~\cite{LiuTan2}): 
%%%%%%%%%%%%%%%%%%%%%%%%%%%%%%%%%%%%%%%%%%%%%%%%%%%%%%%%%%%%%%%%%%%%%%%%%%%%%%%%%%%%%%%%%%%%%%%%%%%%%%%%%%%%%%
\begin{eqnarray} 
F(X,\zeta,t) & = & 
\frac{1}{2 P^+} \left[ {\overline{U}(P',S')}\left( \gamma^+ H^q(X,\zeta,t)+
\frac{i \sigma^{+ \mu} \Delta_\mu}{2M} E^q(X,\zeta,t) \right)U(P,S) \right] \nonumber \\
& = & \sqrt{1-\zeta} \, H^q(X,\zeta,t) -
\frac{1}{4}\frac{\zeta^2}{\sqrt{1-\zeta}} \, 
E^q(X,\zeta,t)  \nonumber \\
& = & \frac{\sqrt{X} \sqrt{X-\zeta}}{1-X}
\int  d^2 {\bf k}_{\perp} \, \rho^q({k^2},k^{\prime \,2}).
\label{matrix1}
\end{eqnarray}
%\begin{eqnarray}
%{\cal M}^N_{ij} = 
%\overline{U}(P^\prime,S) \overline{\Gamma}(k^\prime,P) 
%\, \frac{(\not\!k^\prime+m)}{k^{\prime 2}-m^2} 
%\, \frac{({\not\!k} + m)}{k^2-m^2} \Gamma(k,P) U(P,S), 
%\label{model_N1}
%\end{eqnarray}
%%%%%%%%%%%%%%%%%%%%%%%%%%%%%%%%%%%%%%%%%%%%%%%%%%%%%%%%%%%%%%%%%%%%%%%%%%%%%%%%%%%%%%%%%%%%%%%%%%%%%%%%%%%%%%
In Eq.~(\ref{matrix1}):
\begin{subequations} 
\begin{eqnarray}
\sqrt{1-\zeta} & = & \frac{1}{2P} Tr\{U(P,S) \overline{U}(P^\prime,S^\prime) \gamma^+\} \\
\frac{1}{4} \frac{\zeta^2}{\sqrt{1-\zeta}} & = & 
\frac{1}{2P} Tr\{U(P,S) \overline{U}(P^\prime,S^\prime) \frac{i}{2M}\sigma^{+ \mu}\Delta_\mu\}, 
\end{eqnarray}
\end{subequations}
where the traces over the nucleon spinors are for the {\em same spin}, $S=S^\prime$, case. 
The factors $\sqrt{X} \sqrt{X-\zeta}$,   
are obtained similarly from the traces over the quarks spinors implicit in 
Eq.~(\ref{matrix1}) \cite{LiuTan2}. Having taken care of the spin structure, we then
model $\rho^q \left( k^2, k^{\prime \, 2} \right)$ as: 
\begin{equation}
\rho^q \left( k^2, k^{\prime \, 2} \right) = 
{\cal N} \frac{\phi(k^{\prime\, 2})}{k^{\prime 2}-m_q^2} \frac{\phi(k^2)}{k^2-m_q^2},
\label{rho}
\end{equation}
$m_q$ is the struck quark's mass, $\phi(k^2)$ is a scalar vertex function 
whose form will be specified in Section \ref{sec3}, and ${\cal N}$ is a normalization constant.
 
By using the components in Eqs.~(\ref{kinematics}) we obtain for $H^q(X,\zeta,t)$:
%%%%
\begin{eqnarray}
\displaystyle
 H^q(X,\zeta,t) & = & \int \, \frac{d^2 {\bf k}_\perp}{(1-X)} \,  
\frac{{\cal A} \, \phi(k^2,\lambda) \phi^*(k^{\prime \, 2},\lambda)}{\left[(2 E_k)(E_p-E_k-E_X) \right]
\left[(2 E_{k^\prime})(E_{p^\prime}-E_{k^\prime}-E_X)\right]} \nonumber \\
%%%%%%%%%%
& = & \int d ^2{\bf k}_\perp
\frac{{\cal A} \, \phi(k^2,\lambda) \phi^*(k^{\prime \, 2},\lambda)}{\left(  {\cal M}_0^2(X) - \frac{{\bf k}_\perp^2}{1-X} -m_q^2 \right)   
\left({\cal M}_\zeta^2(X) - \frac{1-\zeta}{1-X}
\left[{\bf k}_\perp - \frac{1-X}{1-\zeta}{\bf \Delta} \right]^2 - m_q^2 \right)}, \nonumber \\
& &
\label{H-model1}
\end{eqnarray}
%%%%%%%%%%%%%
%%%%%%%%%%%%%
where we have rendered explicit the connection between the time ordered diagram in Fig.~\ref{fig2}a
and the covariant expression in Eq.~(\ref{rho}). Furthermore, in Eq.~(\ref{H-model1}):

\begin{equation}
{\cal M}_\zeta^2(X) = (X-\zeta)/(1-\zeta)  M^2 - (X-\zeta)/(1-X)M_X^{q \, 2},
\label{MX}
\end{equation}
and 
\begin{equation}  
{\cal A} = {\cal N} \frac{X}{1-X} \sqrt{\frac{X-\zeta}{X}} \frac{1}{\sqrt{1-\zeta}} 
\end{equation}

The GPD, $E^q(X,\zeta,t)$ is modeled similarly to $H^q$, but imposing a different normalization,
namely: ${\cal N}_E = \kappa_q {\cal N}_H$, (${\cal N}_H \equiv {\cal N}$ in Eq.~(\ref{rho})), where
$\kappa_q$ is the quark's $q$ component of the anomalous magnetic moment. 

The invariant mass of the spectator, $k_X^2 \equiv M_X^{q}$, appearing in Eq.~(\ref{H-model1}) 
through Eq.~(\ref{MX}),  
is a flavor dependent parameter. Both in Eq.~(\ref{H-model1}), and in the results presented 
in Section \ref{sec3} the value of $M_X^{q}$ is considered to be fixed for each configuration. 
%%%
%%% Regge
%%%
However, a spectral distribution in $M_X^{q \, 2} \equiv (P-k)^2$ should in principle be 
introduced for large values of the invariant mass. 
This affects mainly the low $X$ region, {\it i.e.} where $M_X^q$ is large, 
and it has been successfully reproduced in deep inelastic scattering processes
by introducing an $M_X^q$ dependence 
of the spectral function consistent with Regge behavior \cite{BroCloGun}.
The role of $t$-channel exchanges in DVCS and related processes 
was addressed recently in \cite{SzcLon} where the rather extreme point of view was taken  
that GPDs measure mostly the parton content of the reggeons.
A full treatment of this important point is beyond 
the scope of the present work and will be considered in a forthcoming paper.
Here, we introduce directly a $\zeta$ and $t$ dependent Regge motivated term, in addition to 
the ``diquark'' term given by Eq.~(\ref{H-model1}), and we study their relative 
contribution to phenomenology in Section \ref{sec4}. 
%%%%%

In Eq.~(\ref{H-model1}) we have written explicitly the dependence on the size parameter $\lambda$, 
which is similar to the one used in the ``overlap representation'' based models 
(cf. {\it e.g.} the equivalent parameter for the more commonly used ``gaussian form'' discussed in
Ref.~\cite{Bolz}).
We underline that the model considered here is, in fact, consistent with the 
``overlap representation'' derived for DVCS in \cite{BroDie,DieKro2}, but, due to 
the covariance of the vertex function, it differs from constituent quark models.  
%It therefore does not necessarily predict a zero of $H^q$ at $X=\zeta$.
Due to the covariance of the vertex function, 
the spectator model captures two essential features, or ``self-consistency'' 
conditions: {\it i)} the GPDs are not imposed to be zero at the endpoint $X=\zeta$, 
thus allowing
for an imaginary part of the DVCS amplitude. This is also in accordance with 
the experimental observation  
from DVCS experiments at both HERMES \cite{HERMES} and Jefferson Laboratory \cite{Jlab}; 
{\it ii)} the GPDs are continuous at the endpoint $X=\zeta$. 
Calculations similar to the one presented here 
were performed both within QED \cite{BroDie}, and in a simplified version of 
the covariant model with scalar particles in \cite{BroEst}. 
Both cases are however presented as illustrations that are not meant to be quantitatively
compared to data. On the other side, attempts similar to ours to extract GPDs from the data
\cite{DieKro1}, although physically motivated, are leaning towards mathematical forms 
similar to ``PDF-type'' parametrizations. 

This paper's goal is to combine both the essential dynamical aspects described above,
with a fully quantitative analysis that is made possible by the flexibility of our simplified 
model.  
A direct comparison with inclusive experimental data is only possible 
in the $\zeta =0$ region. 
At $\zeta \neq 0$, one needs to include in the analysis the higher moments of GPDs, that are 
$\zeta$-dependent, besides the nucleon form factors. Higher moments are currently available
from lattice QCD \cite{Schierholz} and can be implemented within an extension of our 
analysis to this case.

%%%%%%%%%%%%%%%%%%%%
%%%%%%%%%%%%%%%%%%%% zeta =0
%%%%%%%%%%%%%%%%%%%%
\subsection{A special case: $\zeta =0$}
\label{sec2b}
The case $\zeta =0$ where the momentum transfer is entirely 
transverse, $\Delta^2 \equiv -\Delta_\perp^2$, plays a special
role since processes of type {\bf (b)}, and {\bf (c)} (see Fig.~\ref{fig2})
are suppressed.

The following relations hold (we set $H^q(X,0,t) \equiv H^q(X,t)$):
\begin{equation}
H^q(X,t=0)  =  q(X), 
\label{qpdf}
\end{equation}
where $q(X)$ is the parton distribution for quark ``$q$''. 
The transverse DIS structure function, $F_T(X) \equiv F_1(X)$ 
is given by:
% and $E^q(X,0,t) \equiv E^q(X,t)$. .
\begin{subequations}
\begin{eqnarray}
F_T^p(X) & = & \frac{4}{9} H^u(X,0)+ \frac{1}{9} H^d(X,0)+\frac{1}{9} H^s(X,0)  \\
F_T^n(X) & = & \frac{1}{9} H^u(X,0)+ \frac{4}{9} H^d(X,0)+\frac{1}{9} H^s(X,0),  
\end{eqnarray}
\label{FT}
\end{subequations}
where we implicitly assume the $Q^2$ dependence, and the structure function $F_2$ is obtained
from Callan-Gross's relation: $2X \, F_1(X) = F_2(X)$ 
Furthermore, the following relations:
\begin{subequations}
\begin{eqnarray}
\int_0^1 dX H^q(X,t) & = & F_1^q(t) \\ 
\int_0^1 dX E^q(X,t) & = & F_2^q(t),  
\end{eqnarray}
\label{FF}
\end{subequations}
define the connection with the quark $q$'s contribution to the Dirac and Pauli form factors.
The proton and neutron form factors are obtained as:
\begin{subequations}
\begin{eqnarray}
F_{1(2)}^p(t) & = & \frac{2}{3}F_{1(2)}^u(t) - \frac{1}{3}F_{1(2)}^d(t) + \frac{1}{3}F_{1(2)}^s(t) \\ 
F_{1(2)}^n(t) & = & -\frac{1}{3}F_{1(2)}^u(t)+  \frac{2}{3}F_{1(2)}^d(t) + \frac{1}{3}F_{1(2)}^s(t),
\end{eqnarray}
\label{FF2}
\end{subequations}
where $F_{1(2)}^s(t)$ was found to be consistent with zero \cite{acha}. 
In our analysis we fitted linear combinations
of the integrals of GPDs obtained from Eq.~\ref{FF} to the electric and 
magnetic form factors, for which the experimental data are more readily accessible:
\begin{subequations}
\label{GEGM}
\begin{eqnarray}
G_E^{p(n)}(t) & = & F_1^{p(n)}(t) + \frac{t}{4M^2}F_2^{p(n)}(t)   \\ 
G_M^{p(n)}(t) & = & F_1^{p(n)}(t) + F_2^{p(n)}(t).
\end{eqnarray}
\end{subequations}

Eqs.~(\ref{qpdf},\ref{FT},\ref{FF},\ref{FF2},\ref{GEGM}) define all the constraints used in our fit.
A detailed description of the results of the fit is presented in Section \ref{sec3}.

%%%%%%%%%%%%
%%%%%% SUBSECTION X<ZETA
%%%%%%%%%%%%
\subsection{The $X < \zeta $ region}
\label{sec2c}
%%%%
When $X < \zeta$, the dominating process is the 
one where a quark with 
$X=k^+/P^+$, and transverse momentum,
${\bf k}_\perp$, and an anti-quark
with $\zeta - X = k^{\prime \, +}/P^+ >0$, and 
${\bf k^\prime}_\perp = {\bf \Delta}_\perp - {\bf k}_\perp$ emitted from the initial
proton, undergo the electromagnetic interaction (Fig.~\ref{fig2}b). 
While in the calculation of form factors, and of the $\zeta=0$ GPDs, process
{\bf (b)} is always suppressed, at $\zeta \neq 0$,
it can represent a situation with ``all particles moving forward'' 
so long as $X < \zeta $. This is evident by inspecting the energy denominators 
in this kinematical region that are, in fact, characterized by similar cancellations as 
for process {\bf (a)} at $X>\zeta$ (Eq.~(\ref{H-model1}). 

The physical 
interpretation of this region still presents, however, a few debatable points.  
Within the overlap representation, GPDs are given exclusively by  
the higher Fock states -- a minimum requirement being the ($q\overline{q} qqq$) state.
The latter are not sufficiently constrained by phenomenological studies. 
%The GPDs in the two kinematical regions also satisfy the continuity requirement.  
Recently, higher Fock states were considered in Ref.~\cite{JICR}  within a LC constituent
quark model for the pion GPDs, 
where their contribution was shown to be indeed 
sizable at $X=\zeta$ (the so-called crossover point).
Notice, however, that as $X$ decreases, a large number of Fock 
components would need to be introduced. 
Quantitative calculations were performed both within QED \cite{BroDie}, and in
a scalar model \cite{BroEst}, where it
was shown that   
the sum of the $X>\zeta$ and $X<\zeta$ contributions naturally provides 
a covariant expression, thus satisfying the polynomiality condition. 
This is 
the independence
of the form factor from the parameter $\zeta$, 
which if not observed, would 
signal the presence of an artificial frame dependence \cite{DieKro2}.
An alternative description is where the $q \overline{q}$ pair contribution to GPDs 
is interpreted as $t$-channel exchanges \cite{RadDD} (modulo appropriate color factors \cite{DieKro2}) 
of either a single meson, or a tower of mesons, as recently proposed in \cite{GuzPol}.  

Lacking a uniform picture, and given the important role that will be played 
by GPDs in the $X<\zeta$ region for the interpretation of a number of experiments: from
$\overline{p}p$, and ${\overline p}A$, exclusive reactions \cite{GSI}, to deep inelastic
pion production and other semi-inclusive experiments at forthcoming 
facilities, we propose a strategy that can provide 
further guidance in this region.  

Starting from the accurate parametrization of the $\zeta=0$ case
presented in this paper, and obtained directly from experimental data,  
we subsequently study the constraints for the $\zeta > 0$ case. These are provided,
on one side,
by the higher order Mellin moments of $H$ and $E$'s  that govern the behavior with 
$\zeta$ of the GPDs.  Moments of order $n \leq 3$ can be obtained from lattice 
calculations {\cite{zan,LHPC}. 
%Their lattice calculations can be extended to all $n$ values according to the 
%prescription of Schierholz \cite{Schierholz}.  
On the other side, we notice that the $X>\zeta$ region is dominated by the 
same ``valence type'' configurations as for the $\zeta=0$ case 
described by Eq.~(\ref{H-model1}), and it can be therefore 
obtained by extending our $\zeta=0$ parametrization to this kinematics using the 
same constraints.  
In this approach, polynomiality is imposed at every step, within a ``bottom up'' type of approach, 
rather than  
the ``top down'' method implicit in both the double distributions \cite{RadDD}.
In Section \ref{sec4} we illustrate the type of information that can be obtained 
from lattice results in the specific cases of $\zeta=0$. 
Information on the $X < \zeta$ behavior using  higher moments, 
combined with the experimentally constrained 
$X>\zeta$ behavior, is extracted according to a deconvolution procedure 
described in detail in a forthcoming paper \cite{part2}.

%%%%%%%%%%%%%%%%%%%%%%%% SECTION III: figures 3 to 10
%%%%%%%%%%%%%%%%%%%%%%%%
\section{Results}
\label{sec3}
We now present our quantitative determination of the unpolarized GPDs, $H$ and $E$, obtained
at $\zeta =0$ using all available data on the proton and neutron electromagnetic
form factors, as well as the valence quarks distributions from DIS measurements. 
Our fit is obtained at a low scale, $Q^2_o \approx 0.1$ GeV$^2$, 
%HH
%and $\Lambda_{QCD} = 0.250$ GeV, 
in line with the approach of Ref.~\cite{Gluck} where it is assumed that at a low scale the 
nucleon consists mostly of valence quarks, the bulk of the gluon and anti-quark distributions being
generated dynamically.  
  
We reiterate that, although direct measurements of DVCS cannot be currently implemented, 
our procedure produces effectively a ``parametrization'', in that parameter-dependent
physically motivated functional forms are fitted to data. The goodness of the fit is tested
by means of a $\chi^2$, whose values, along with the parameter errors have been quantitatively 
evaluated and are given below. We, of course, support the future usage of DVCS data because they 
are more directly linked to GPDs \cite{Camacho}, and we are actively 
considering their implementation \cite{part2}. 
The amount of data and their kinematical range is however too limited at present to  
provide sensibly more stringent constraints. We would also like to add that parametrization shapes
do represent a possible bias in the present analysis as well as in 
any type of fitting (see {\it e.g.} discussion in 
Ref.~\cite{Pumplin}), for instance
the shape of the gluon distribution functions has been oscillating through the years 
between ``valence-like'' to hard-peaked at $x \rightarrow 0$. 
The problem of the initial bias can be attacked in a similar way for GPDs, by tuning in possible new
shapes as constraints from new sets of data become available, allowing for more refined fitting.  
 
Starting from Eq.~(\ref{H-model1}), with the inclusion of the Regge term discussed in Section 
\ref{sec2a} we obtained two slightly different
forms that are both constrained by current experimental data:

\vspace{0.5cm}
\noindent {\bf Set I} 
\begin{eqnarray}
H^I(X,t) & = & G_{M_{X}^I}^{\lambda^I}(X,t) \,  
 X^{-\alpha^I -\beta_1^I (1-X)^{p_1^I} t}
\label{param1_H}
\\
E^I(X,t) & = & \kappa \, G_{M_X^I}^{\lambda^I}(X,t) \,
X^{-\alpha^I -\beta_2^I (1-X)^{p_2^I} t}
\label{param1_E}
\end{eqnarray}

\noindent {\bf Set II} 
\begin{eqnarray}
H^{II}(X,t)& = & G_{M_X^{II}}^{\lambda^{II}}(X,t) \,
X^{-\alpha^{II} - \beta_1^{II}  (1-X)^{p_1^{II}} t}
\label{param2_H}
\\
E^{II}(X,t) & = & G_{\widetilde{M}_X^{II}}^{\widetilde{\lambda}^{II}}(X,t) \,
X^{-\widetilde{\alpha}^{II} - \beta_2^{II}  (1-X)^{p_2^{II}} t}
\label{param2_E}
\end{eqnarray}
%%%
All parameters except for $p_1$ and $p_2$ are flavor dependent; we omit, however,
the ``q'' symbol (unless specifically needed) 
for ease of presentation.
The function $G$ has the same form for both parametrizations, I and II:
% and 
%it is obtained using a model, defined below, for the vertex function $\phi$ 
%(Eq.~(\ref{H-model1})): 
\begin{eqnarray}
G_{M_X}^{\lambda} (X,t) = && 
{\cal N} \frac{X}{1-X} \int d^2{\bf k}_\perp \frac{\phi(k^2,\lambda)}{D(X,{\bf k}_\perp)}
\frac{\phi({k^{\prime \, 2},\lambda)}}{D(X,{ \bf k}_\perp +(1-X){\bf \Delta}_\perp)},    
\label{gkaava}
\end{eqnarray}
where
%HH:
\begin{equation}
D(X,{\bf k}_\perp) \equiv  k^2 - m^2,
\end{equation}
and
\begin{eqnarray}
k^2 & = & X M^2 - \frac{X}{1-X} M_X^{2}  - \frac{{\bf k}_\perp^2}{1-X} \\
k^{\prime \, 2} & = & XM^2 - \frac{X}{1-X} M_X^{2} - \frac{({\bf k}_\perp - (1-X)\Delta)^2}{1-X},
\end{eqnarray}
$m$ being the struck quark mass, and $M$, the proton mass.
The normalization factor includes the nucleon-quark-diquark coupling, and it
is  set to ${\cal N} = 1$ GeV$^6$.  
Eq.~(\ref{gkaava}), was obtained 
using the following Dirac structure for the vertex in Eq.~(\ref{H-model1}): 
\begin{equation}
\Gamma(k,P)_{\alpha\beta} = 
\sum_\lambda u_\alpha(k,\lambda)\overline{U}_\beta(P,\lambda) \, \phi(k^2,\lambda^2),   
\end{equation}
where \cite{MelSchTho}
%HH:
%%%
\footnote{With this choice of diquark form factor, one ensures that the 
value of the quark mass, or equivalently the position of the 
pole in the quark propagator, does not play a dynamical role in the model.
The values of the quark masses are consequently not determined in our fit.}    
%%%
\begin{eqnarray}
\phi(k^2,\lambda) = && \frac{k^2-m^2}{\vert k^2-\lambda^2\vert^2 }.
\label{phi}
\end{eqnarray}
Finally, the $u$ and $d$ quarks contributions to the anomalous magnetic moments are:
\begin{equation}
\kappa \equiv \kappa^q= \left\{ \begin{array}{ll}
\kappa^d = 2.03,& \; {\rm for} \; q=d
\\
\kappa^u/2 = 1.67/2, & \; {\rm for}\; q=u
\end{array} \right. .
\label{kappaq}
\end{equation}
A few comments are in order:

\vspace{0.3cm}
\noindent
{\it i)} At present it is important to provide a parametrization that allows one 
to address a richer phenomenology, including the interplay of coordinate and momentum space
observables. 
The spectator model used here is ideal because despite its simplicity, it 
has proven to be sufficiently flexible to 
describe (and predict) the main features of a number of distribution and fragmentation
functions in the intermediate and large $X$ regions, as well as the unintegrated PDFs  
\cite{MeyMul,Jakob:1997wg}.  
%%%%%%% flavor dependence
The spectator system can be a scalar or a spin 1 vector, thus allowing us to access both the $u$ and
$d$ quark distributions. In  Ref.~\cite{MeyMul} it was shown that the value of 
the spectator mass in the two cases is crucial in shaping the parametrizations for $u$ and
$d$ quarks respectively. Here we have let also the mass parameter $\lambda$ be flavor dependent. 

\vspace{0.3cm}
\noindent {\it ii}) Similarly to the case of DIS structure functions 
the spectator model is not able to reproduce quantitatively the very small
$X$ behavior of the GPDs (this proplem is again present  
in the very small $t$ behavior of the nucleon form factors and GPDs).
This mismatch is not very visible in the results of Refs.~\cite{MelSchTho,MeyMul,Jakob:1997wg}
because of the linear scale used in the plots; 
it is, however, responsible for a violation of the baryon number sum rule that becomes
particularly important in GPD parametrizations since one needs to achieve
a precise agreement with the nucleon form factors as well.  
We introduced therefore a ``Regge-type'' term 
multiplying the spectator model function $G_{M_X}^{\lambda}$ in 
Eqs.~(\ref{param1_H},\ref{param1_E},\ref{param2_H},\ref{param2_E}).
A similar behavior was considered in the ``profile functions'' of Refs.~\cite{DieKro1,VandH1}. 
%Set 2 are flexible enough to get an excellent small $t$ fit without violating 
%the baryonumber sum rule and Eq.~\ref{mmom}. 
However, our procedure is distinctively different  (as discussed also in our results below) 
because in our case a simultaneous fit to both the PDFs from DIS and to the nucleon form factors
is performed. In Refs.~\cite{DieKro1,VandH1} the PDF limit (Eq.~(\ref{qpdf})) 
is trivially satisfied, whereas the form factors and the additional constraints from the expected 
Regge behavior are subsequently used to define the GPDs shape.   

\vspace{0.3cm}
\noindent {\it iii)} We considered the two variants shown above, in order to 
estimate the sensitivity to different procedures as also described in {\it ii)}. 
Set I and II differ in 
the determination of $E$, that is in principle unrelated to the forward PDFs, and therefore
less constrained by the data. 

\subsection{Results of fit from nucleon form factors and PDFs}
\label{sec3a}
%HH
%All available experimental data on the nucleon
%form factors were implemented in the fit:
The experimental data on the nucleon
form factors implemented in the fit are:
$G_{E}^p$ \cite{GEP_exp},
$G_{M}^p$ \cite{GMP_exp}, 
$G_{E}^n$ \cite{GEN_exp}, 
$G_{M}^n$ \cite{GMN_exp}
and $G_{E}^p/G_{M}^p$ \cite{GEP_GMP}.
The data selection is the same that was used in Ref.~\cite{Kelly:2004hm}, where 
for $-t > 1$ GeV$^2$ only the measurements based on polarization transfer techniques
were considered, while the Rosenbluth separation ones were discarded.
%HH
In the fitting procedure
% each one
%of the form factors was weighted by the number of corresponding data points.
%Note that
all of the form factor data enter simultaneously in the parametrizations 
for $H^q$ and $E^q$, respectively. 
This is at variance with implementing data
on the $F_1$ and $F_2$ form factors, which also require extrapolations from 
different data sets.
By fitting directly to the electric and magnetic form factors we obtained a 
more precise determination
since no data manipulation is necessary. In particular,
an accurate description of the
low $t$ region is important in view of future comparisons with the lattice 
determinations \cite{part2}.  

The $\chi^2$ per number of data points in each data set, as well as for the total number 
of data points, is listed in Table \ref{table_khi2} for parametrization Sets I and II.
The comparison with form factor data is shown in Fig.~\ref{fig3} for the proton, and in 
Fig.~\ref{fig4} for the neutron. The ratio $G_M^p/G_E^p$ is shown in Fig.~\ref{fig5}.
%HH
To check to what extent our parametrization is dominated by the data from 
$G_M^p$, we also repeated the fit by weighting each form factor by the number 
of corresponding data points. The effect for the listed values of $\chi^2$ was
less than $2\%$.
%%%
\begin{table}[h]
\center
\begin{tabular}{|c|c|c|c|}
\hline
\hline
Data Set  & $\chi^2/{\mbox N_{\rm data}}$ Set 1 & $\chi^2/{\mbox N_{\rm data}}$ Set 2  & Data Points \\ 
\hline
\hline
$G_{E_p}$ &  1.049 &  0.963 & 33 \\ \hline
$G_{M_p}$& 1.194 & 1.220 & 75 \\ \hline
$G_{E_p}/G_{M_p}$  & 0.689 & 0.569  & 20 \\ \hline
$G_{E_n}$ & 0.808 & 1.059  & 25 \\ \hline
$G_{M_n}$ & 2.068 & 1.286 & 24 \\ \hline
TOTAL  & 1.174   &  1.085 & 177 \\ 
\hline
\hline
\end{tabular}
\caption{The $\chi^2/N_{\rm data}$ of the different nucleon form factors obtained from 
Set I and Set II. $N_{data}$ is the number of 
data points available for each set of form factor data.
}
\label{table_khi2}
\end{table}
%%%

%%%% Procedure
The parameters in the $t \rightarrow 0$ limit 
were determined by fitting to the LO set of Alekhin PDFs \cite{Alekhin:2002fv} 
within the range 
$10^{-5} \le x \le 0.8$ and  $4 \le Q^2 \le 240$ GeV$^2$, and imposing the
baryon number and momentum sum rule such that: 
\begin{eqnarray}
&&\int^1_0 dX  u(X,Q_0^2)  =  2 \\
&&\int^1_0 dX d(X,Q_0^2)  =  1 \\
&&\int^1_0 dX X \left[ u (X,Q_0^2) + d (X,Q_0^2) \right]  =  1.
\label{pdfmomsr}
\end{eqnarray}
%for including 
%them would emphasize the large-$x$ region where our model fails to reproduce 
%the shape of the PDFs.

The parameters involved in this step, $M_{X}^q$, $\lambda^q$ and $\alpha^q$, $q=u,d$, obtained
at an initial scale $Q_o^2$ ($Q^2_o = 0.094$ GeV$^2$), are listed in Table \ref{table_DIS}. 
Notice that: {\it i)} they are the same for both Sets I and II; {\it ii)} in Set I they are by 
definition the same
for the functions $H$ and  $E$ (see Eqs.~(\ref{param1_H},\ref{param1_E})).
%HH
The parameters in the PDCD evolution were chosen as in CTEQ6L1 \cite{CTEQ}.

%%%
\begin{table}[h]
\center
\begin{tabular}{|c|c|c|c|}
\hline
\hline
Flavor  &  $M_X$ (GeV) & $\lambda$ (GeV) & $\alpha$ \\ 
\hline
\hline
u &  0.4972  &  0.9728 & 1.2261 \\ \hline
d &  0.7918  &  0.9214 & 1.0433 \\ \hline
\hline
\hline
\end{tabular}
\caption{Parameters fixing the shape of $H^q$, $q=u,d$, at $t=0$. The parameters are the
same for both Set I and Set II. Moreover, they also define the $t=0$ limit for $E^q$ in Set I, 
as it can be seen from the definitions given in Eqs.~(\ref{param1_H},\ref{param1_E}).}  
\label{table_DIS}
\end{table}
%%%%

Similar results can be in principle obtained from other current PDF parametrizations
\cite{CTEQ,MRST}, however the valence contributions from Ref.~\cite{Alekhin:2002fv} tend
to more readily agree with the shape given in Eqs.~(\ref{param1_H},\ref{param2_H}).     
The experimental data on DIS structure functions were not used directly, 
because our parametrization does not include an ansatz for the sea quarks. Therefore
it is not possible to reproduce within this context
the low $x$ behavior of the DIS structure function $F_2(X,Q^2)$. 
For the same reason, the estimated errors of Alekhin's PDFs were not used in the fit. 
This aspect is beyond the scope of the present analysis, and 
will be improved in future work on by extending our model to include 
sea quarks \cite{part2}.
%using another kind of vertex function the vertex-function,
%for example varying the exponent  $\alpha$, 
It should be noticed that, as in similar approaches \cite{MeyMul,MelSchTho,BroEst},
higher-order effects might be important especially considering 
the low value of the initial scale, $Q_0^2$, resulting from the requirement that only valence quarks
contribute to the momentum sum rule (Eq.~(\ref{pdfmomsr})). 
However, on one side some of the previous evaluations \cite{MelSchTho,Gluck} do not seem to 
find quantitatively large higher order effects, on the other, the fact that $Q_0^2$ is a 
parameter in our model, determined itself by fitting the valence PDF to the (parametrization
of the) data \cite{Alekhin:2002fv} in the large $Q^2$ regime, lends self-consistency to
our procedure (see also \cite{Gluck,SchThoLon}). In order to gauge the effects
of perturbatuive evolution, a dedicated study of NLO is being considered in upcoming future work.   

With the parameters given in Table II the values of the baryon number sum rules for $u$ and $d$ are:   
1.9998 and  1.0003, respectively. The momentum at the initial scale 
adds up to  1.0016. 
For Set II we imposed the additional normalization condition:
\begin{equation}
\int^1_0 dX E_q(X,t=0) = \kappa^q,
\label{mmom}
\end{equation}  
the experimental values of $\kappa^q$, $q=u,d$ being given in 
Eq.~(\ref{kappaq}), 
while our fitted values were 1.6715 and  -2.0309, for $u$ and $d$ 
respectively. 
The comparison with the PDF parametrization of Ref.~\cite{Alekhin:2002fv} 
is shown in Fig.~\ref{fig6}.

%%%%%
The parameters $\beta_1$, $\beta_2$, ${p_1}$ and ${p_2}$, in Set I, and 
all parameters defining $E$ in Set II (Eq.~({\ref{param2_E})),
were fitted  to the nucleon electric and magnetic form factors, Eqs.~(\ref{GEGM}),
with the values of $M_X^q$, $\lambda^q$, and $\alpha^q$ fixed as in Table \ref{table_DIS}.

In Table \ref{table_FF1} we list the values of the parameters for Set I, with 
their corresponding 1-$\sigma$ errors 
%HH
(which for 6 parameters corresponds to
$\Delta\chi^2=7.04$).
%HH We reiterate that we resolved the problem pointed 
%HH out in Ref.~\cite{DieKro1}, of possible biases
%HH in the form factor fits due to an uneven distribution of data in either the %HH low or high $t$ 
%HH regions by using weighted $\chi^2$'s as shown in Table \ref{table_khi2}.   
The error band on the form factors resulting from the 1-$\sigma$ errors on the parameters
is displayed in Figs.~\ref{fig3}, \ref{fig4} and \ref{fig5}. 

%%%
\begin{table}[h]
\center
\begin{tabular}{|c|c|c|c|c|}
\hline
\hline
Flavor  &  $\beta_1$ (GeV$^{-2}$) & $\beta_2$ (GeV$^{-2}$) & $p_1$ & $p_2$ \\ 
\hline
\hline
u &  1.9263 $\pm$ 0.0439 & 3.0792 $\pm$ 0.1318 & 0.720 $\pm$ 0.028 & 0.528 $\pm$ 0.031 \\ \hline
d &  1.5707 $\pm$ 0.0368 & 1.4316 $\pm$ 0.0440 & 0.720 $\pm$ 0.028 & 0.528 $\pm$ 0.031 \\ \hline
\hline
\hline
\end{tabular}
\caption{Parameters fixing the $t$ behavior of the GPD forms given in Set I
(Eqs.~(\ref{param1_H},\ref{param1_E})). The subscript $1(2)$ is for function 
$H^q(E^q)$, for each flavor.}
\label{table_FF1}
\end{table}
%%%

In Table \ref{table_FF2} we show the values of $\beta_1$, $\beta_2$, ${p_1}$ and ${p_2}$
for Set II. We do not present their corresponding errors because 
due to the additional number of paramemeters in this variant of the parametrization, the
fit tends to be over-determined, and therefore not completely quantitative. 
This problem can be in principle circumvented either 
with an increased flow of new data from DVCS experiments, or by reducing the number of 
parameters by keeping some of their values fixed. While these strategies can be
addressed in the future, we consider variant II of our fit as an indicative measure
of the model dependence of the $E$ component.    

%%%
\begin{table}[h]
\center
\begin{tabular}{|c|c|c|c|c|}
\hline
\hline
Flavor  &  $\beta_1$ (GeV$^{-2}$) & $\beta_2$ (GeV$^{-2}$) & $p_1$ & $p_2$ \\ 
\hline
\hline
u & 1.9567  & 0.1767  & 0.742 & 0.270 \\ \hline
d & 1.5896  & 3.2866  & 0.742 & 0.270  \\ \hline
\hline
\hline
\end{tabular}
\caption{Parameters fixing the $t$ behavior of the GPD forms given in Set II 
(Eqs.~(\ref{param2_H},\ref{param2_E})). The subscript $1(2)$ is for function 
$H^q(E^q)$, $q=u,d$.}
\label{table_FF2}
\end{table}

Additional parameters in Eqs.~(\ref{param2_H},\ref{param2_E}) are given by: 
$\widetilde{M}_{X_u}^{II}=  1.5780 \, {\rm GeV}$,
$\widetilde{M}_{X_d}^{II}=  0.3902 \, {\rm GeV}$,
$\widetilde{\lambda}_u^{II} = 0.2678\, {\rm GeV}$, 
$\widetilde{\lambda}_d^{II} = 0.9589\, {\rm GeV}$, 
$\widetilde{\alpha}_u^{II} =  0.005381$, $\widetilde{\alpha}_d^{II} =  0.7501$.
Despite the larger number of parameters, the overall agreement with the data is not 
significantly different from Set I: 
the small $t$ shape of the $G_{M_n}$ data seems is better reproduced, 
due to the increased flexibility, whereas with Set I one
obtains a slightly better description at large $t$. 

\subsection{$\zeta =0$ GPDs for $u$ and $d$ quarks}
\label{sec3b}  
Our results for the GPDs are presented in Figs.~\ref{fig7},\ref{fig8}, \ref{fig9}, \ref{fig10}, and 
\ref{fig11}, respectively.
In Fig.~\ref{fig7} we show  $H_u$ and $H_d$ plotted vs. $X$ at different values of $t$
for the parameters in Set I (results for Set II are within the $2 \%$ 
error band shown in the figure.
Plots for $E_u$ and  $-E_d$ are shown in Fig.~\ref{fig8} for Set I. In Fig.~\ref{fig9}
we display results for $E_u$ and $E_d$, for both Set I and Set II. Differently from  
$H$, in this case there is a much bigger discrepancy in the shape of the curves at the initial scale,
$Q_0^2$, due to the fact that the constraint from the PDFs is missing in this case. 
As is will shown later on, however, PQCD evolution reduces substantially this discrepancy. 

%%%%%%%%%%%%%%% figs 10 and 11 (Regge)
In Fig.~\ref{fig10} we show the separate contributions to $H_u$ 
of the diquark, $G_{M_X}^{\lambda}$, Eq.~(\ref{gkaava}),
and Regge terms, $R = X^{-\alpha - \beta (1-X)^{p_1} t}$, respectively, at the fixed values of 
$t=-0.08$ GeV$^2$, and $t=-1.8$ GeV$^2$. 
%%%%%%%%%%%
From the figure it appears clearly that the form of the GPDs is determined by 
the diquark shape, with an ``envelope'' provided by the Regge term.
 
%%% ADD REGGE
We reiterate that we did not attempt at guiding 
the values for the parameters in our Regge motivated term, based on results from
soft hadron interactions phenomenology, as done instead in Ref.~\cite{DieKro1} and Ref.~\cite{VandH1},
respectively. Therefore, our final parametrization, 
does not depend entirely on the Regge or diquark
behaviors in either the low or large $X$ regions, but on a mixture of both. 

In Fig.~\ref{fig11} and Fig.~\ref{fig12} we show the importance
of the Regge term, in a quantitative fit of PDFs. 
In Fig.~\ref{fig11} the $u$-valence distribution, $u_v(X,Q^2)$ 
for our full parametrization, including both the 
Regge and quark-diquark term (see Eqs.~(\ref{param1_H}), and (\ref{param2_H}), is compared 
to both the PDF fit of 
Ref.~\cite{Alekhin:2002fv}, and to fits performed by excluding the Regge term, at $Q^2= 5$ GeV$^2$. 
The quality of the ``non-Regge'' type
fits is similar to those performed {\it e.g.} in \cite{Jakob:1997wg} and \cite{MelSchTho}, however 
it is clear 
that the diquark fit alone does not provide hard enough distributions at low $X$. This
result is also independent
from whether one relaxes the constraints from the baryon number and momentum sum rules.     
We reiterate that the term, $X^{-\alpha}$ which is essential in obtaining the argument at low $X$, is 
also necessary in order to obtain the correct value of the baryon sum rule. To test this
we performed weighted fits to the low 
$X$ region without the Regge term. By forcing the model to fit accurately at low $X$ resulted 
in a large mismatch at larger $X$. Similar results are obtained from other variants of the diquark
model. We conclude that the diquark model is not apt to reproduce the low $X$ behavior, and the baryon
sum rule which is in turn fundamental for a quantitative fit of GPDs from the nucleon form factors.      
Fig.~\ref{fig12} emphasizes the low $X$ region on a logarithmic scale. It is shown in particular, 
how the final result
(full curve) is determined by both contributions of our Regge-motivated term (thick dotted curve),
and the diquark model term (dashed curve). Our fit curve is shown to be in relatively good agreement  
with the term: 
$X^{-\alpha^\prime - \beta^\prime t}$, where $\alpha^\prime$ and $\beta^\prime$ take values 
consistent with Regge determinations of soft hadronic cross sections (shaded area in the figure).  
The interplay between the diquark model and the Regge term explains why the values of 
the ``Regge'' parameters in our fit differ sensibly from the results from hadronic cross sections, 
nevertheless giving an accurate description of the low $X$ and low $t$ regions.   

%%%%%%%% Comparisons 
Both recent parametrizations from Ref.~\cite{DieKro1} and Ref.~\cite{VandH1}, are compared with ours
for $H^{u(d)}$ in Fig.~\ref{fig13}, and in Fig.~\ref{fig14}, respectively. A similar comparison for 
$E^{u(d)}$ is performed in Figs.~\ref{fig15} and \ref{fig16}.
The results presented in the figure were obtained by evolving to Leading Order (LO) 
the parametrizations at the initial 
scale $Q_0^2 \approx 0.1$ GeV$^2$ to the values of $Q^2=1$ GeV$^2$, and $Q^2=4$ GeV$^2$, 
where the parametrizations from 
\cite{DieKro1} and \cite{VandH1} were respectively given. We reiterate that in principle higher order 
effects are important especially considering the low values of the scale, and that PQCD evolution
for GPDs involves many more subtleties that already appear at NLO 
(see {\it e.g.} \cite{Die_rev} and references therein). Lacking however, any knowledge from experiment,
it is still important to take into account the effects of PQCD evolution, as indicators of the general 
trend followed by GPDs.      

At low $t$ we notice a very good agreement in both $H^u$ and $H^d$ among all three approaches, 
essentially because the GPDs tend to the forward limit constrained by PDFs, 
whereas a disagreement appears in in the large $t$ behavior. This is clearly an effect of perturbative 
evolution that sets in earlier in our case with respect to Refs.~\cite{DieKro1,VandH1}, and that 
does not conserve the form of the initial function, as also noticed in \cite{DieKro1}.
One can gain insight on this point, as first observed in the preliminary study of Ref.~\cite{LiuTan1}, 
by plotting the quantity: 
%%%%
%%%% <X>
%%%%
\begin{equation}
\langle X^q (t)\rangle = \frac{\int_0^1 dX \, X \, H^q(X,t)}{\int_0^1 dX \, H^q(X,t)} \equiv
\frac{1}{F_1^q(t)} \, \int_0^1 dX \, X \, H^q(X,t), 
\label{xave}  
\end{equation}
representing the average value of $X$ contributing respectively, to the $u$ and $d$ 
components of the nucleon form factors, $F_1^u$ and $F_1^d$ (Eq.~(\ref{FF2})). 
In Fig.~\ref{fig17} we plot $\langle X^q (t)\rangle$ at $Q^2 \equiv Q_0^2$. 
The contribution to the proton form factor is obtained as:
\begin{equation}
\langle X^p (t)\rangle = \frac{2}{3} \frac{F_1^u}{F_1^p} \langle X^u (t)\rangle  - 
\frac{1}{3} \frac{F_1^d}{F_1^p} \langle X^d (t)\rangle.
\end{equation}
Notice that the total momentum carried by the valence $u$ and $d$ quarks is instead given by: 
\begin{equation}
\langle X \rangle = 2 \langle X^u(0) \rangle + \langle X^d(0) \rangle. 
\label{momsr}
\end{equation}
From Fig.~\ref{fig17} one can see that at $Q^2= Q^2_0$, $\langle X \rangle = 1$, 
{\it i.e.} all the momentum
is carried by the valence quarks. PQCD evolution implies that:
\begin{equation}
\langle X^q(t,Q^2) \rangle = \langle X^q(t,Q_0^2) \rangle 
\left[\alpha_s(Q^2)/\alpha_s(Q_0^2) \right]^{d_2},
\label{qcdmomsr}
\end{equation}
where $\alpha_s$ is the strong coupling constant, and the anomalous dimension yields 
$d_2 = 0.4267$. One can therefore observe that although larger values of $X$ tend to dominate
the form factor as $t$ increases, the effect of PQCD evolution reduces the large 
$X$ components by a $Q^2$-dependent shift. This effect explains the discrepancies in the curves
at $-t= 5$ GeV$^2$ in Figs.~\ref{fig12},\ref{fig13},\ref{fig14},\ref{fig15}.     

A similar behavior is observed for $E^{u(d)}$, although all parametrizations tend to quantitatively 
differ also at low $t$, since they are not constrained by the PDFs.

%%%%%%%%%%%%%%%%%%%%%%%% SECTION IV: figures 16 to 20%
%%%%%%%%%%%%%%%%%%%%%%%%
\section{Phenomenology}
\label{sec4}
With the results of a precise fit at hand, we can address some issues
in the phenomenology of GPDs at zero skewness.

\subsection{Coordinate Space Observables}
\label{sec4a}
A most interesting aspect is the relation with  
the nucleon Impact Parameter dependent PDFs (IPPDFs), $q(X,{\bf b})$, related via
Fourier transformation to $H^q(X,t \equiv - {\bf \Delta}_\perp^2)$ \cite{Bur}:
\begin{eqnarray}
q(X,{\bf b}) & = & \int d^2 \Delta \, e^{i {\bf b \cdot \Delta}}  H^q(X,t),  
\label{qxb}
\end{eqnarray}
$q(X,{\bf b})$ is the probability of finding a quark in the 
proton carrying momentum fraction $X$, at impact parameter ${\bf b}$. 

The quark's average impact parameter can be derived as:
\begin{equation}
\langle b_q^2 (X)\rangle  =  \frac{\int d^2{\bf b} \, \, q(X,{\bf b}) \, b^2}{\int d^2{\bf b} \, \, q(X,{\bf b})}
= 4 \, \frac{\partial}{\partial t} \log H^q(X,t) \Bigg|_{t=0},  %\eqpt
\label{impact_b}
\end{equation}
from which an average ``interparton distance'' \cite{Soper} can be defined as:
\begin{eqnarray}
\langle y_q^2(X) \rangle =  \frac{\langle b_q^2 (X)\rangle}{(1-X)^2}.
\label{yx}
\end{eqnarray}

Similarly, $E^q$ can be interpreted in terms of a distribution function 
in a transversely polarized target, through:
\begin{eqnarray}
q^X(X,{\bf b}) & = & q(X,{\bf b}) - \frac{b_y}{M} \frac{\partial}{\partial {\bf b}^2} 
\int d^2 \Delta \, e^{i {\bf b \cdot \Delta}}  E^q(X,t),  
\label{qXxb}
\end{eqnarray}
where polarization is along the $x$ axis and $b_y$ is the component of ${\bf b}$ along the 
$y$ axis. $q^X(X,{\bf b})$ measures the probability of finding a quark carrying momentum fraction 
$X$ in the transversely 
polarized proton, at impact parameter ${\bf b}$.  
The average shift in the quark's distance along the $y$-axis for polarization along 
the $x$-axis direction is obtained from Eq.~(\ref{qXxb}) as \cite{Bur}:
\begin{equation}
\langle b_q^y (X)\rangle  =  
\frac{\int d^2{\bf b} \, \, q^X(X,{\bf b}) \, b_y}{\int d^2{\bf b} \, \, q^X(X,{\bf b})} 
= \frac{1}{2M} \frac{E^q(X,0)}{H^q(X,0)}. 
\label{impact_by}
\end{equation}
The shift relative to the spectator quarks is obtained analogously to Eq.~(\ref{yx}) as: 
\begin{eqnarray}
\langle s_q (X) \rangle =  \frac{\langle b_q^y (X)\rangle}{1-X}.
\label{sx}
\end{eqnarray}
The average interparton distances, and the transverse shifts are shown in Fig.~\ref{fig18}, and 
Fig.~\ref{fig19}, respectively.

%%%%%%%%%%%%%%%%%% Radii  
In our approach the radii are a result of the fitting procedure rather than an additional constraint
as in Refs.~\cite{DieKro1,VandH1}. 
We studied the role of the Regge-type and diquark terms from 
Eq.~(\ref{H-model1}). Because of the factorized form,  $\langle y_q^2(X) \rangle$ can be expressed in fact
as the sum of the two terms. We find a larger than intuitively 
expected contribution of the diquark term to 
both the $d$ and $u$ quarks interparton distances. 
Notice that the interparton distances are subject to PQCD evolution. In Fig.~\ref{fig18} 
we display results at our initial low scale, $Q_0^2$. The effect of evolution is shown by plotting
the total interparton distance -- including both Regge and diquarks terms -- at $Q^2 = 4$ GeV$^2$.
One can see that the interparton distances tend to decrease 
with $Q^2$, as the term $\partial H/\partial t$ in Eq.~(\ref{impact_b}) evolves more steeply than $H$. 

The total radius squared of the nucleon is obtained by considering:
\begin{equation}
\langle r_1^{q \, 2} \rangle  =  \frac{\int dX \, q(X) \langle b_q^2(X)  \rangle}{F_1^q(t)}\Bigg|_{t=0} 
= 4 \, \frac{\partial}{\partial t} \log F_1^q(t) \Bigg|_{t=0},  
\label{b1}
\end{equation}
and using isospin symmetry:
\begin{subequations}
\begin{eqnarray}
\langle r_1^{p \, 2} \rangle & = &  \frac{4}{3} \langle r_1^{u \, 2} \rangle 
- \frac{1}{3}\langle r_1^{d \, 2} \rangle \\
\langle r_1^{n \, 2} \rangle & = &  -\frac{2}{3} \langle r_1^{u \, 2} \rangle 
+ \frac{2}{3}\langle r_1^{d \, 2} \rangle 
\end{eqnarray}
\end{subequations}
The charge radii are given by: 
\begin{subequations}
\begin{eqnarray}
\langle r_E^{p \, 2} \rangle & = &  \langle r_1^{p \, 2} \rangle + \frac{3}{2}\frac{\kappa_p}{M^2} \\
\langle r_E^{n \, 2} \rangle & = &  \langle r_1^{p \, 2} \rangle + \frac{3}{2}\frac{\kappa_n}{M^2},
\end{eqnarray}
\end{subequations}
We find: $\langle r_1^{u \, 2} \rangle = 0.654$ fm$^2$,  
$\langle r_1^{d \, 2} \rangle = 0.666$ fm$^2$, $\langle r_1^{p \, 2} \rangle = 0.650$ fm$^2$,   
$\langle r_1^{n \, 2} \rangle = $ 0.0078 fm$^2$. The calculated values for the total charge radii are:
$\langle r_E^{p \, 2} \rangle = 0.76$ fm$^2$  and  
$\langle r_E^{n \, 2} \rangle = -0.118$ fm$^2$, in agreement with the experimental results.
Therefore our results for the fit parameters $\alpha_q \approx 1 - 1.2$ in Eq.~(\ref{H-model1}) are  
in line with the constraints on the Regge parameters studied in Ref.~\cite{VandH1}.  
 
The transverse shift, $s_q$, (Fig.~\ref{fig19}) constitutes in principle a test on the GPD $E^q$.
Our results, plotted at the initial scale, 
show a marked difference between Set I and Set II. One should keep in mind however, that 
PQCD evolution largely diminishes the discrepancy. 

%%%%
%%%% k_\perp
%%%%
\subsection{Intrinsic Transverse Momentum}
\label{sec4b}
An observable related to transversity that can be accessed witihin our model, 
is the partons' average intrinsic transverse momentum, $\langle k_\perp^2(X) \rangle_q$.
Notice that  ${\bf k}_\perp$, although not Fourier conjugate to ${\bf b}$,   
it can be related to $q(X,{\bf b})$ as shown in Ref.~\cite{LiuTan1}.
%%%%%%%%%
\footnote{Such relation is of general validity, unrelated to non-relativistic many body theory
as quoted in \protect\cite{BelRad}.}
%%%%%%%%% 
In other words, 
${\bf k}_\perp$ is not directly observable in DVCS type processes. However, it can be 
evaluated using the same input to Eq.~(\ref{H-model1}), as:
\begin{subequations}
\begin{eqnarray}
\langle k_\perp^2(X) \rangle_q & = & \frac{\int d^2{\bf k}_\perp \, \mid \phi(k^2,\lambda^2) \mid^2 
{\bf k}_\perp^2}{\int d^2{\bf k} \, \mid\phi(k^2,\lambda^2) \mid^2 }, \\
\langle k_\perp^2(X) \rangle & \equiv & \langle k_\perp^2(X) \rangle_p = 
\frac{4}{9} \langle k_\perp^2(X) \rangle_u + \frac{1}{9} \langle k_\perp^2(X) \rangle_d
\label{kperp}
\end{eqnarray}
\end{subequations}
In Fig.~\ref{fig20} we show $\langle k_\perp^2 (X) \rangle$ from our model.
In order to assess the range of possible variations for this observable, we 
compare our evaluation with the values extracted from Refs.~\cite{DieKro1} and \cite{VandH1}, 
by assuming a gaussian $k_\perp$ dependence of the vertex functions that could originate such
parametrizations. 
We also compare with the hypothesis originally advanced by Burkardt \cite{Burk2} on 
the form of the combined $X$ and $t$ dependences of the gaussian's exponent 
(or the profile function in \cite{DieKro1,VandH1}). Finally, we compare 
with the values used in Semi-Inclusive DIS (SIDIS) parametrizations \cite{Jakob:1997wg,anselmino}.
The exploratory work presented here is aimed at defining a few guidelines 
for future quantitative studies of the connection between GPDs and SIDIS reactions \cite{Bur,Hagler}.
%%%%
%%%% What X and k_\perps are the form factors dominated by?
%%%%
In Fig.~\ref{fig21} we show the 
ratios: 
\begin{eqnarray}
R_{k_{max}}^{1(2)}(t) & = & F_{1(2)}(t, k_{max})/F_{1(2)}(t) \\
R_{X_{max}}^{1(2)}(t) & = & F_{1(2)}(t, X_{max})/F_{1(2)}(t) 
\end{eqnarray}
in order to determine
what $k_\perp$,  and $X$ components the form factors are dominated by. 
The numerators were obtained by setting the upper limit of 
integration in Eq.~(\ref{FF}) to different values of $k_\perp \equiv k_{max}$, and 
$X \equiv X_{max}$, respectively. 
It can be clearly seen that for all values of $t$ the form factor 
ratio, $R_{k_{max}}^{(1)}(t)$,  is saturated by setting $k_{max} \approx 1$ GeV 
(for $F_2$, even larger values of $k_\perp$ seem to be important). 
This is in turn an indication
of a semi-hard distribution, as opposed to the soft gaussian forms used elsewhere.
On the other side, the ratio $R_{X_{max}}^{1(2)}(t)$ 
clearly shows the coupling between the $X \rightarrow 1$
behavior of the GPDs and the large $t$ behavior of the form factors.

\subsection{Implementation of Lattice QCD Results}
\label{sec4c}
Lattice QCD provides the only ``model independent'' constraints that are necessary 
to parametrize GPDs at $\zeta \neq 0$. It is, however, important to illustrate the type of information
that can be obtained on GPDs starting from the $\zeta=0$ case. The $\zeta \neq 0$ case 
requires in addition, a more involved ``deconvolution'' procedure from the first three 
moments that will be described in a following dedicated  paper \cite{part2}.   

Current lattice results can reproduce the dipole fall-off of the form factors up to 
$-t \approx 3$ GeV$^2$. However, even after performing a linear extrapolation to low values of the 
pion mass, such calculations overshoot the experimental results \cite{zan,LHPC,DieKro1}.  
As a result, predictions from the lattice are at the moment characterized by a rather 
large uncertainty. This becomes problematic especially for the higher moments of GPDs where 
no comparison with experimental results can be made. 
In order to evaluate the impact of such uncertainty on possible extractions of GPDs, 
we performed a test based on a prescription proposed in \cite{Schierholz} by Schierholz, 
according to which the value of the dipole mass, $\Lambda_n$, appearing in:
\footnote{We reiterate that Eq.~(\ref{lattice_mom}) is not of general validity but a result 
of fits to the lattice calculations of \cite{zan,LHPC}.}   
%%%
\begin{equation}
M_n(t) = \int_0^1 dX H(X,t) X^{n-1} = M_n(0) \frac{1}{\left( 1+\frac{t}{\Lambda_n^2} \right)^2}, 
\label{lattice_mom}
\end{equation}
can be first extracted from the lattice evaluations for $n \leq 3$, and subsequently 
extended to all $n$ values by performing a fit based on a Regge 
motivated ansatz of the type:
\begin{equation}
\sqrt{\Lambda_n^2} = \frac{n-\alpha_0}{\alpha_v}.
\label{dip_lattice}
\end{equation}
(in Eq.~(\ref{lattice_mom}) $H = H^u-H^d \equiv H^{u-d}$ evaluated at $Q^2 = 4$ GeV$^2$). 
The $n$-dependence of the moments described above allows one to perform the anti-Mellin
transform, thus obtaining GPDs, analytically. 
Results using Eqs.~(\ref{lattice_mom},\ref{dip_lattice}), 
are compared with the evaluations from our analysis at $Q^2 = 4$ GeV$^2$, in Fig.~\ref{fig22}. 
The band in the figure 
includes the estimate of the lattice error. One can observe a good agreement with both our 
results and other current parametrizations for the $H^u$ at $t \lesssim 2$ GeV$^2$. The agreement
however, deteriorates at lower $t$ for the $d$-quark. 
We therefore conclude that within the range of $t$ shown in the figure, 
it is acceptable to use current lattice results in order to extend our analysis of 
the extraction of GPDs at $\zeta \neq 0$.  

\section{Conclusions and Outlook}
In this paper we proposed a fully quantitative physically motivated parametrization of generalized parton 
distributions that is constructed directly from a covariant model for the quark-nucleon 
scattering amplitude. Therefore, we do not implement a specific form for the forward limit given 
by the parton distribution functions of DIS, but we obtain that limit from 
fitting directly to the valence contribution to DIS data. This allows us in particular to better study the 
role of Regge-type exchanges, that are disengaged, in our case, from the specific form 
of parton distributions.
The other constraints defining our parametrization at zero skewness are 
provided by the electric and nucleon form factor data.
 
The advantages of this approach are that on one side, using the same initial formalism, 
we can predict additional quantities such as the the unintegrated, $k_\perp$-dependent, parton 
distribution functions. 
This degree of flexibility is desirable
in view of both future interpretations of both coordinate space observables, and of 
transversity. 
Furthermore, analyzing directly the vertex structure of the scattering amplitude 
allows us to easily extended our predictions to the non-zero skewness case.  
This can be, in fact, obtained by extending the range of the kinematical variables in our expressions
for the quark scattering dominated region, and by considering separately 
the $X < \zeta$ region where a quark-antiquark pair from the
initial proton participates in the scattering process.
Additional constraints need however to be provided by the higher, $\zeta$-dependent, 
moments of the GPDs. These are in principle available from recent lattice calculations but 
are necessarily fraught with uncertainties. Here were able to assess the impact
of such ambiguities on the extraction of GPDs. We conclude that current extrapolated lattice 
values can be used for $t \lesssim 2$ GeV$^2$. 
A detailed description of the $\zeta$ dependent parametrization is given in a forthcoming paper.

With this type of parametrization in hand we can on one side provide predictions 
for both recent \cite{Ava,Camacho} and future DVCS measurements at Jefferson Lab. On 
the other, our approach is geared towards providing a practical and more flexible method to 
reconstruct generalized parton distributions from their first few moments.

\acknowledgments
We thank S. J. Brodsky, F. Llanes-Estrada and P. Kroll for useful comments.
This work is supported by the U.S. Department
of Energy grant no. DE-FG02-01ER41200 and NSF grant no.0426971. 

\clearpage
%%%%%%% Appendix 
\appendix
\section{Connection between different sets of kinematical variables} 
 
By defining \cite{Ji1}: $x  =  (k^+ + k^{\prime +})/(P^+ + P^{\prime +})$, 
$\xi =  \Delta^+/(P^+ + P^{\prime +})$, and $t$, the following mappings with 
the kinematical variables defined in the paper obtain:
The regions $- \xi < x < \xi $ and $\xi < x <1 $  are mapped into 
$0 < X < \zeta $ and $\zeta < X <1$ respectively, via:
\begin{subequations}
\begin{eqnarray}
X & = & \frac{x+\xi}{1+\xi} \\
\zeta & = & \frac{2\xi}{1+\xi},
\label{kin1}
\end{eqnarray}
\end{subequations}
while the region $ -1 < x < - \xi $ maps into $\zeta < X <1$, via :
\begin{subequations}
\begin{eqnarray}
X & = & \frac{-x+\xi}{1+\xi} \\
\zeta & = & \frac{2\xi}{1+\xi} 
\label{kin2}
\end{eqnarray}
\end{subequations}

\section{Analytic Expressions for the Nucleon's  Radius}
\label{analytic_r}
We give the analytic expressions for $\langle b_q^2(X) \rangle$, obtained from the 
factorized form in Eq.~(\ref{param1_H}):
\begin{eqnarray}
\langle b_q^2(X) \rangle & = & 4 \, \frac{\partial}{\partial t} \log H^q(X,t) \Bigg|_{t=0} 
\nonumber \\
& = & -\frac{1}{2\Delta} \left[ \frac{\partial \ln G_{M_X}^\lambda}{\partial \Delta} + 
\frac{\partial \ln R}{\partial \Delta} \right] \Bigg|_{\Delta=0} 
\\ \nonumber 
& = & \left( \langle b_q^2(X) \rangle_G + \langle b_q^2(X) \rangle_R \right)
\frac{1}{q(X)} \\ \nonumber 
& = &  \frac{3}{5} \frac{1-X}{(-M^2 + \frac{X}{1-X}M_X^{q \, 2} - \lambda_q^2)} 
- \beta_1 (1 -X)^{p_1} \, \log X  
\end{eqnarray}
with $\Delta = \sqrt{-t}$, $G_{M_X}^\lambda$ given in Eq.(\ref{gkaava}), 
and $R= X^{-\alpha -\beta_1 (1-X)^{p_1} t}$.
 
%%%%%%%%%%%%%%%%%%%%%%%%%%%%%%%%%%%%%%%%%%%%%%%%%%%%%%%%%%%%%%%%%%%%%%%%%%%%%%%%%%%%%%%%%%%%%%%%%%
%%%%%%%%%%%%%%%%%%

\clearpage

\newpage
%%%
%%% FIGURES
%%%%%%%%%%%%%%%%%%%%%%%%%%%%%%%%%%%%%%%%%%%%%%%%%%%%%%%%%%%%%%%%%%%%%%%%%%%%%%%%%%%%%%%%%%%%%% 
%%%%%%%%%%%%%%%%%%%%%%%%%%%%%%%%%%%%%%%%%%%%%%%%%%%%%%%%%%%%%%%%%%%%%%%%%%%%%%%%%%%%%%%%%%%%%%
%%%
%%% FIGURE 1
%%%
\begin{figure}
\includegraphics[width=12.cm]{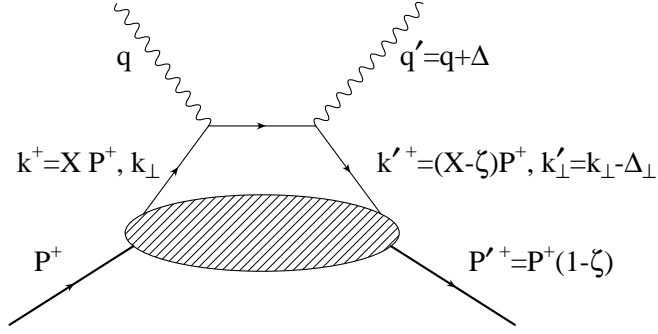}
\caption{Amplitude for DVCS at leading order in $Q^2$. The light-cone coordinates for the active 
quarks and nucleons are explicitly written.} 
\label{fig1}
\end{figure}

\newpage
%%%%%%%%%%%%%%%%%%%%%%%%%%%%%%%%%%%%%%%%%%%%%%%%%%%%%%%%%%%%%%%%%%%%%%%%%%%%%%%%%%%%%%%%%%%%%% 
%%%%%%%%%%%%%%%%%%%%%%%%%%%%%%%%%%%%%%%%%%%%%%%%%%%%%%%%%%%%%%%%%%%%%%%%%%%%%%%%%%%%%%%%%%%%%%
%%%
%%% FIGURE 2
%%%
%\begin{widetext}
\begin{figure}
%{8.5cm}
%\hspace{2cm}
\includegraphics[width=5.cm]{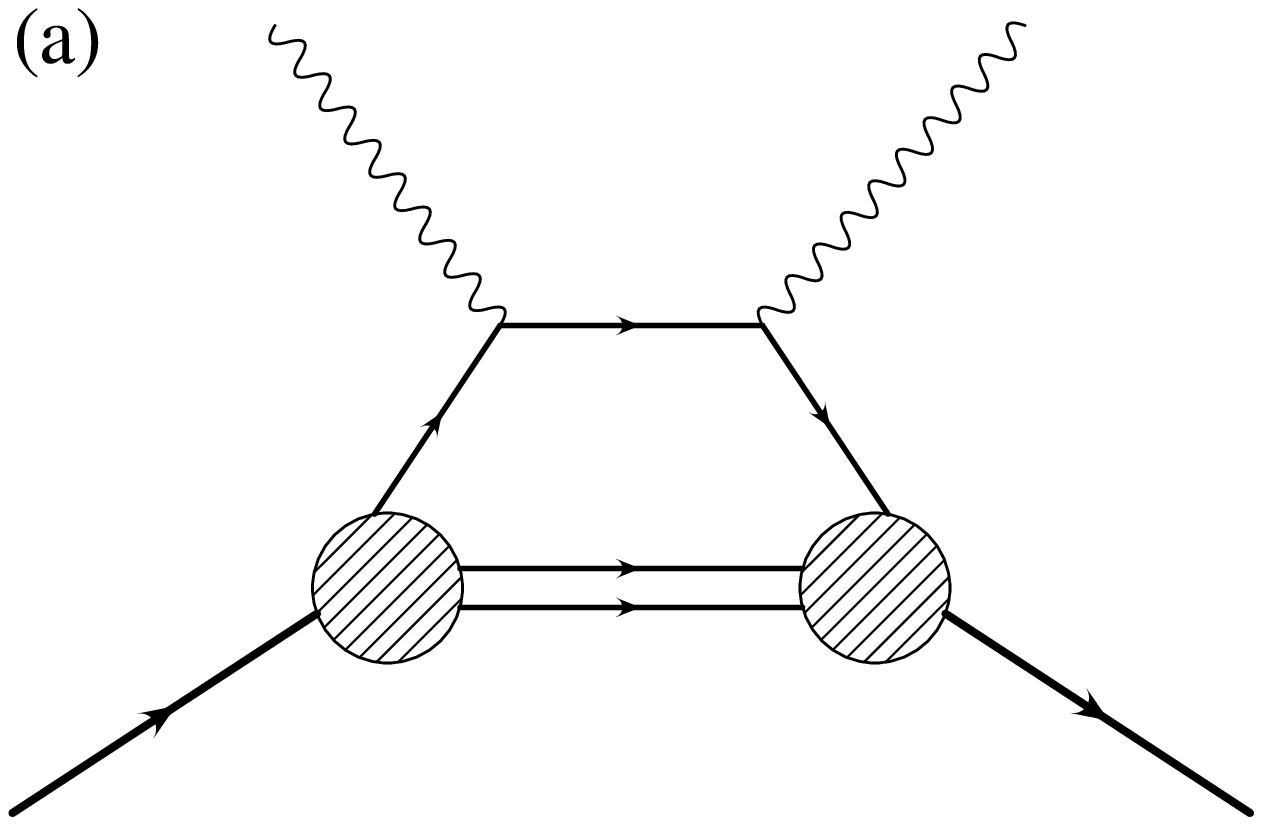}%
\hspace{0.7cm}
\vspace{0.5cm}%
\includegraphics[width=6cm]{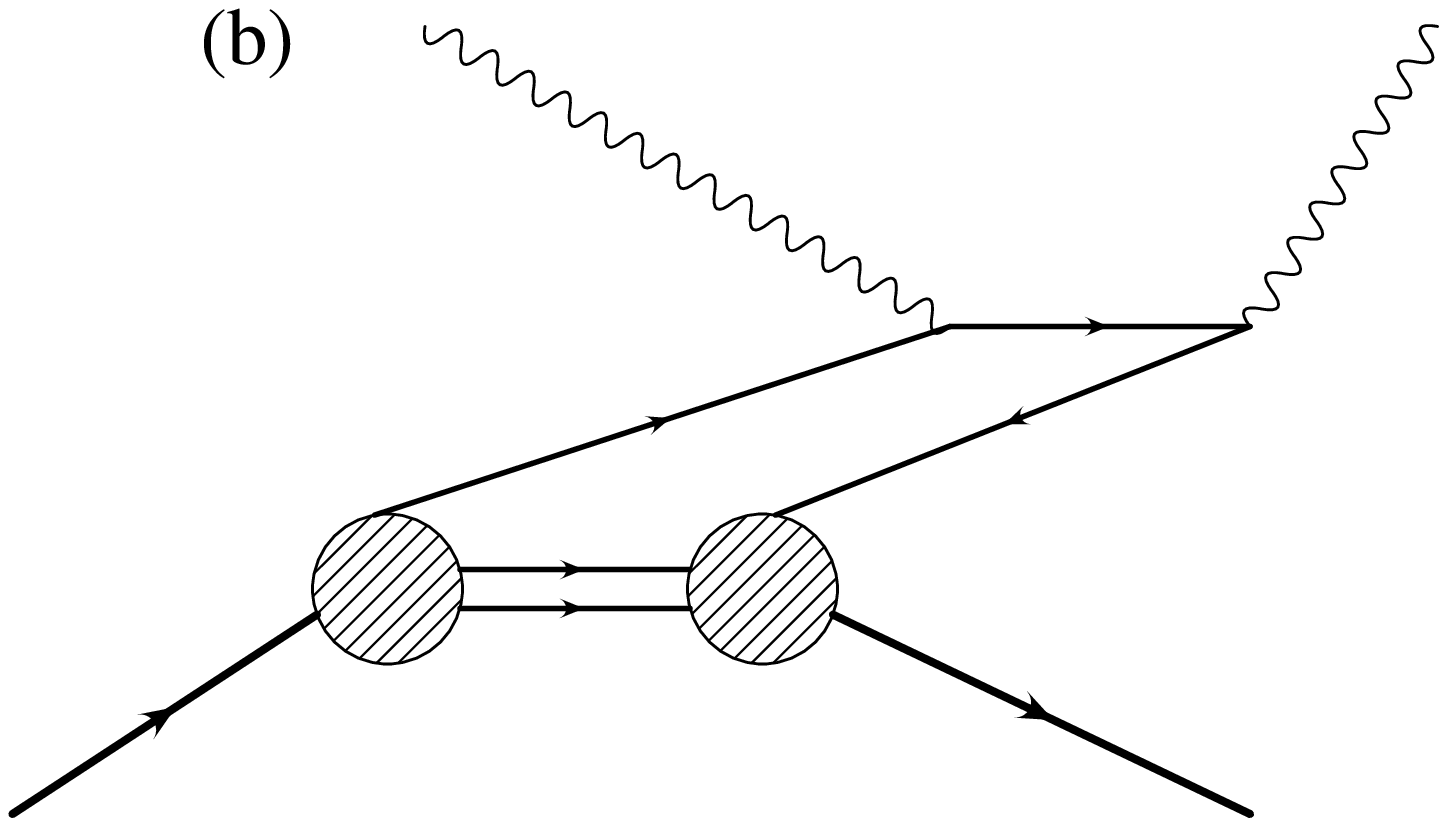}
\includegraphics[width=6.5cm]{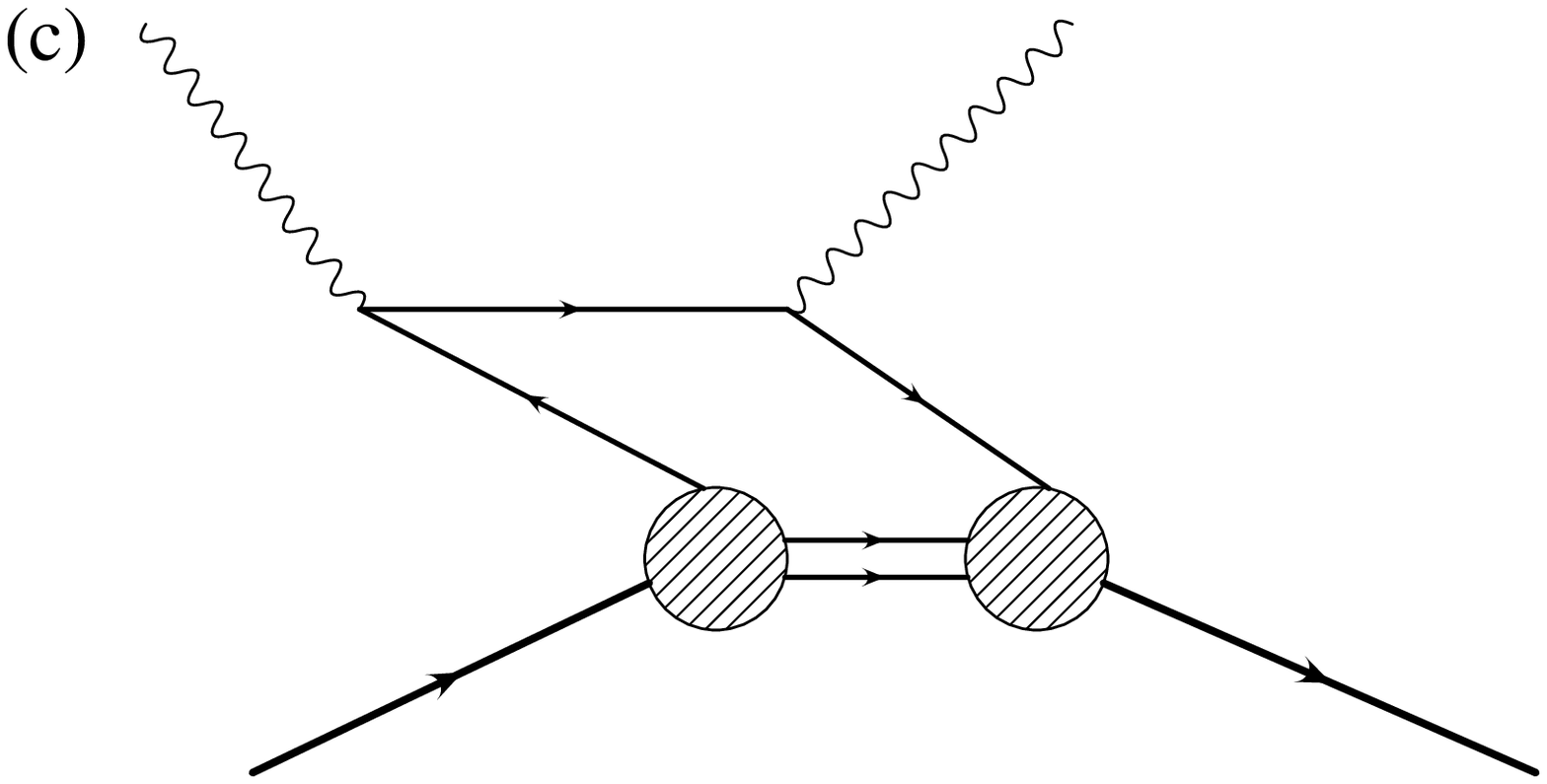}
\caption{Time ordered diagrams for DVCS: {\bf (a)} dominant contribution in $X>\zeta$ region;
{\bf (b)} a $q \overline{q}$ pair is first produced from the nucleon and subsequently 
interacts with the photons. This process dominates the $X<\zeta$ region; {\bf (c)} the initial
photon splits into a $q \overline{q}$ pair that interacts with the hadronic system. 
The crossed-terms where 
two of the particles in the same class are switched, are not shown in the figure.}
\label{fig2}
\end{figure}
%\end{widetext}

%%%%%%%%%%%%%%%%%%%%%%%%%%%%%%%%%%%%%%%%%%%%%%%%%%%%%%%%%%%%%%%%%%%%%%%%%%%%%%%%%%%%%%%%%%%%%% 
%%%%%%%%%%%%%%%%%%%%%%%%%%%%%%%%%%%%%%%%%%%%%%%%%%%%%%%%%%%%%%%%%%%%%%%%%%%%%%%%%%%%%%%%%%%%%% 
%%%%%%%%%%%%%%%%%%%%%%%%%%%%%%%%%%%%%%%%%%%%%%%%%%%%%%%%%%%%%%%%%%%%%%%%%%%%%%%%%%%%%%%%%%%%%% 
%%%%%%%%%%%%%%%%%%%%%%%%%%%%%%%%%%%%%%%%%%%%%%%%%%%%%%%%%%%%%%%%%%%%%%%%%%%%%%%%%%%%%%%%%%%%%% 
%%%
%%% FIGURE 3
%%%
\begin{figure}
%{8.5cm}
%\hspace{2cm}
\includegraphics[width=8.cm]{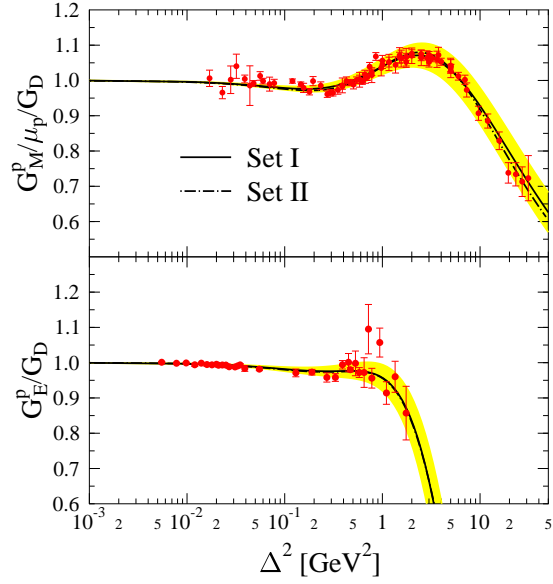}
\caption{The proton magnetic and electric form factors, $G_M^p$ and $G_E^p$, respectively, 
divided by the dipole, $G_D=1/(1+Q^2/0.71 \, {\rm GeV}^2)^2$, plotted vs. $\Delta^2$. 
Experimental data from  
\protect\cite{GEP_exp} ($G_{E_p}$);
\protect\cite{GMP_exp} ($G_{M_p}$ ). The full line was obtained 
using parametrization I, Eqs.~(\ref{param1_H},\ref{param1_E}); the dot-dashed line corresponds to parametrization II,
Eqs.~(\ref{param2_H},\ref{param2_E}).}  
\label{fig3}
\end{figure}
%%%%%%%%%%%%%%%%%%%%%%%%%%%%%%%%%%%%%%%%%%%%%%%%%%%%%%%%%%%%%%%%%%%%%%%%%%%%%%%%%%%%%%%%%%%%%% 
%%%%%%%%%%%%%%%%%%%%%%%%%%%%%%%%%%%%%%%%%%%%%%%%%%%%%%%%%%%%%%%%%%%%%%%%%%%%%%%%%%%%%%%%%%%%%% 
%%%
%%% FIGURE 4
%%%
\begin{figure}
%{8.5cm}
\hspace{2cm}
\includegraphics[width=8.cm]{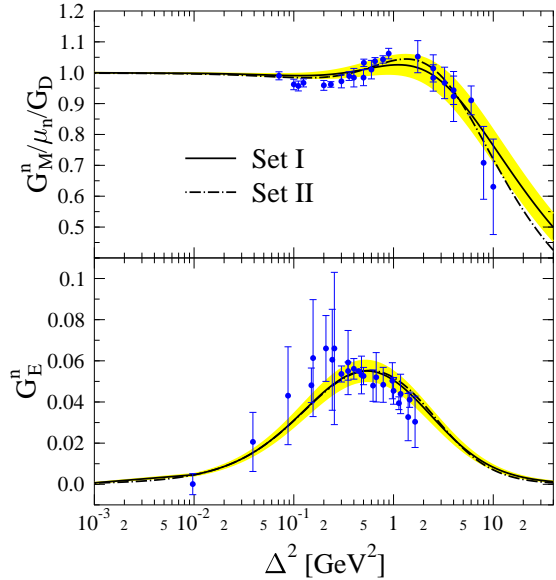}
\caption{The neutron magnetic form factor, $G_M^n$, divided by the dipole, 
$G_D=1/(1+\Delta^2/0.71 \, {\rm GeV}^2)^2$,
and electric form factor, $G_E^n$, respectively, plotted vs. $\Delta^2$. 
Experimental data from  
\protect\cite{GMN_exp}($G_{M_n}$) and \protect\cite{GEN_exp} ($G_E^n$). 
Notation as in Fig.~\ref{fig3}.} 
\label{fig4}
\end{figure}
%%%%%%%%%%%%%%%%%%%%%%%%%%%%%%%%%%%%%%%%%%%%%%%%%%%%%%%%%%%%%%%%%%%%%%%%%%%%%%%%%%%%%%%%%%%%%% 
%%%%%%%%%%%%%%%%%%%%%%%%%%%%%%%%%%%%%%%%%%%%%%%%%%%%%%%%%%%%%%%%%%%%%%%%%%%%%%%%%%%%%%%%%%%%%% 
%%%%%%%%%%%%%%%%%%%%%%%%%%%%%%%%%%%%%%%%%%%%%%%%%%%%%%%%%%%%%%%%%%%%%%%%%%%%%%%%%%%%%%%%%%%%%% 
%%%%%%%%%%%%%%%%%%%%%%%%%%%%%%%%%%%%%%%%%%%%%%%%%%%%%%%%%%%%%%%%%%%%%%%%%%%%%%%%%%%%%%%%%%%%%% 
%%%
%%% FIGURE 5
%%%
\clearpage
\begin{figure}
%{8.5cm}
\hspace{2cm}
\includegraphics[width=8.cm]{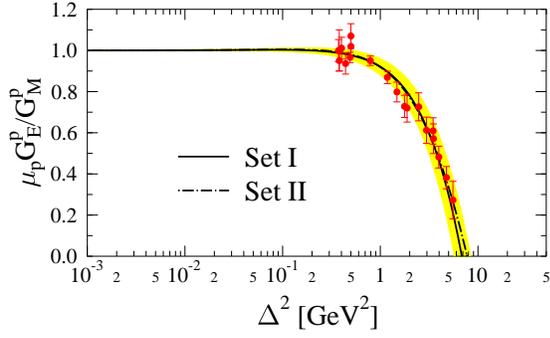}
\caption{The ratio of proton electric and magnetic form factors, $\mu_p G_M^p(t)/G_E^p(t)$. 
Experimental data from \protect\cite{GEP_GMP}.
Notation as in Fig.~\ref{fig3}.}  
\label{fig5}
\end{figure}
%%%
%%% FIGURE 6 
%%%
\begin{figure}
%{8.5cm}
\includegraphics[width=8.cm]{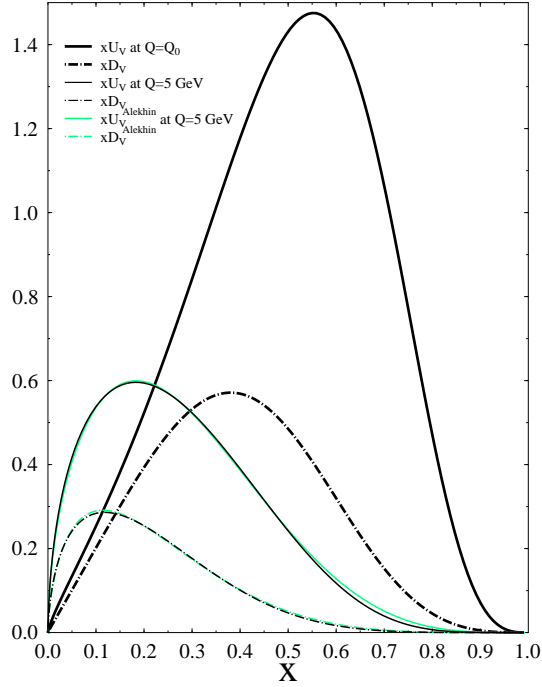}
\caption{(color online) Parton distribution functions, $X u_v(X)$ and $X d_v(X)$ plotted vs. $X$, 
at the initial scale, $Q_0=0.3 $ GeV, and at $Q = 5$ GeV. The parton distributions 
parameters, including the value of the initial scale were obtained directly from our fit. 
The fitted parameters are: $M_X$, $\lambda$, $\alpha$ 
for both parametrization I and II. Our results are shown together with the 
LO set of Alekhin PDFs \cite{Alekhin:2002fv} used in the fit.}  
\label{fig6}
\end{figure}

\newpage
%%%%%%%%%%%%%%%%%%%%%%%%%%%%%%%%%%%%%%%%%%%%%%%%%%%%%%%%%%%%%%%%%%%%%%%%%%%%%%%%%%%%%%%%%%%%%% 
%%%%%%%%%%%%%%%%%%%%%%%%%%%%%%%%%%%%%%%%%%%%%%%%%%%%%%%%%%%%%%%%%%%%%%%%%%%%%%%%%%%%%%%%%%%%%% 
%%%
%%% FIGURE 7
%%%
\begin{figure}
%{8.5cm}
\includegraphics[width=8.cm]{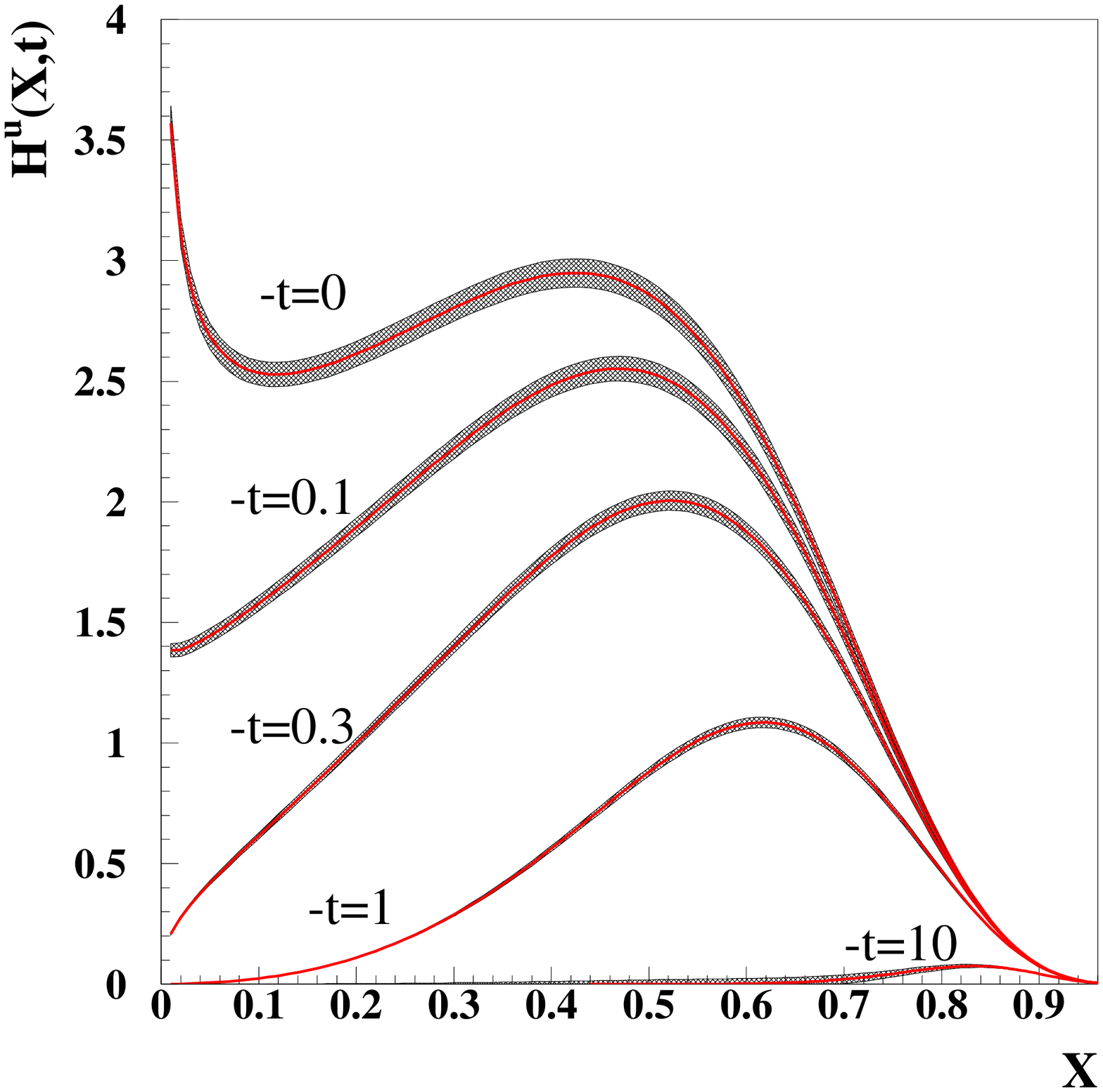}
\includegraphics[width=8.cm]{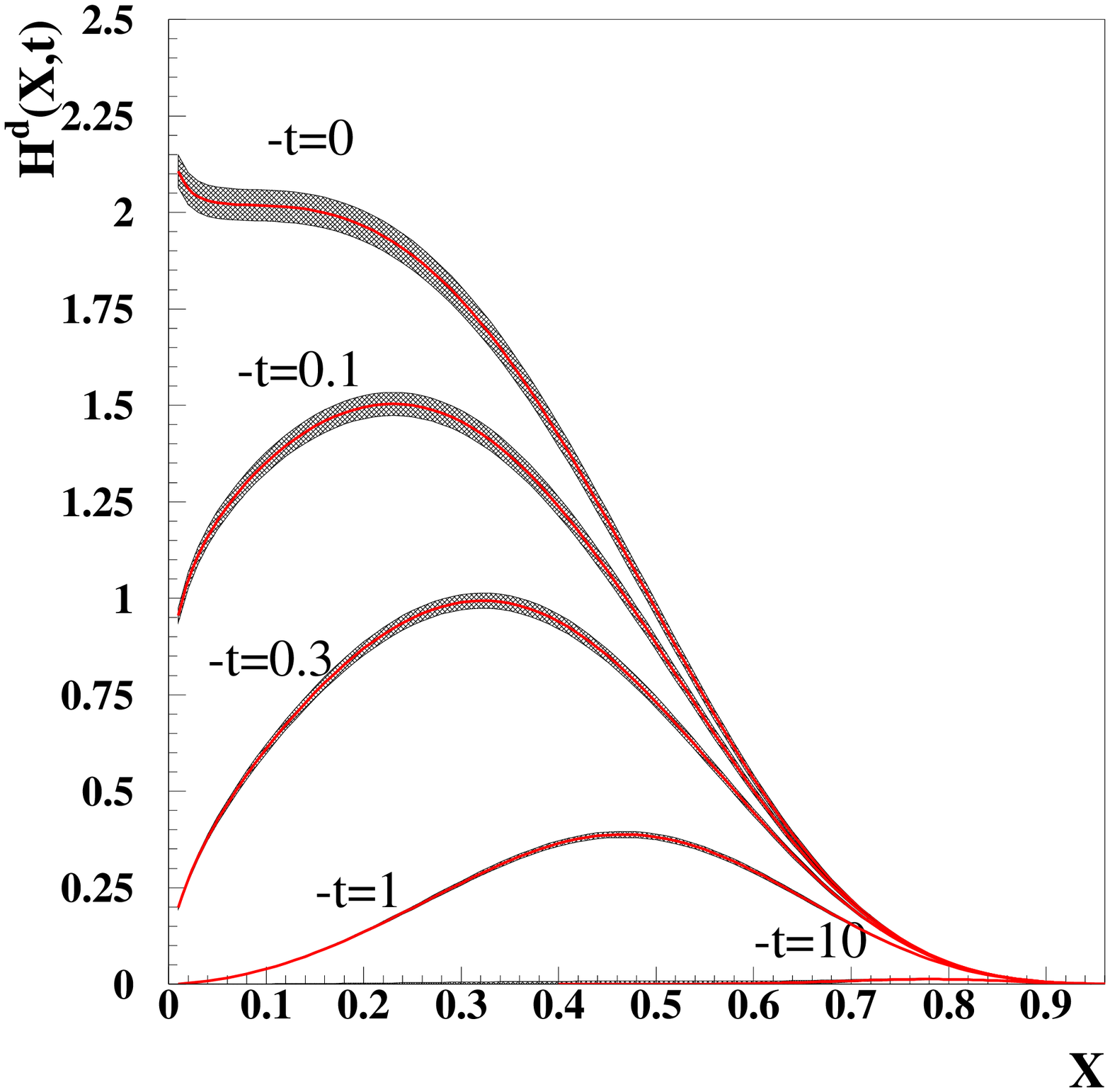}

\caption{The generalized parton distributions, $H_u$ (left panel) and $H_d$ (right panel) 
obtained from Parametrization I, defined by Eq.~(\ref{param1_H}), 
plotted vs. $X$ at $-t=0,0.1,0.3,1, 10$ GeV$^2$, respectively, at the initial scale
$Q_0^2 = 0.09$ GeV$^2$. 
Results for Parametrization II, Eq.~(\ref{param2_H}) are very similar. 
They are included within the $2 \%$ band displayed in the figure}   
\label{fig7}
\end{figure}
%%%%%%%%%%%%%%%%%%%%%%%%%%%%%%%%%%%%%%%%%%%%%%%%%%%%%%%%%%%%%%%%%%%%%%%%%%%%%%%%%%%%%%%%%%%%%% 

%%%%%%%%%%%%%%%%%%%%%%%%%%%%%%%%%%%%%%%%%%%%%%%%%%%%%%%%%%%%%%%%%%%%%%%%%%%%%%%%%%%%%%%%%%%%%% 

\newpage
%%%
%%% FIGURE 8
%%%
\begin{figure}
%{8.5cm}
\includegraphics[width=8.cm]{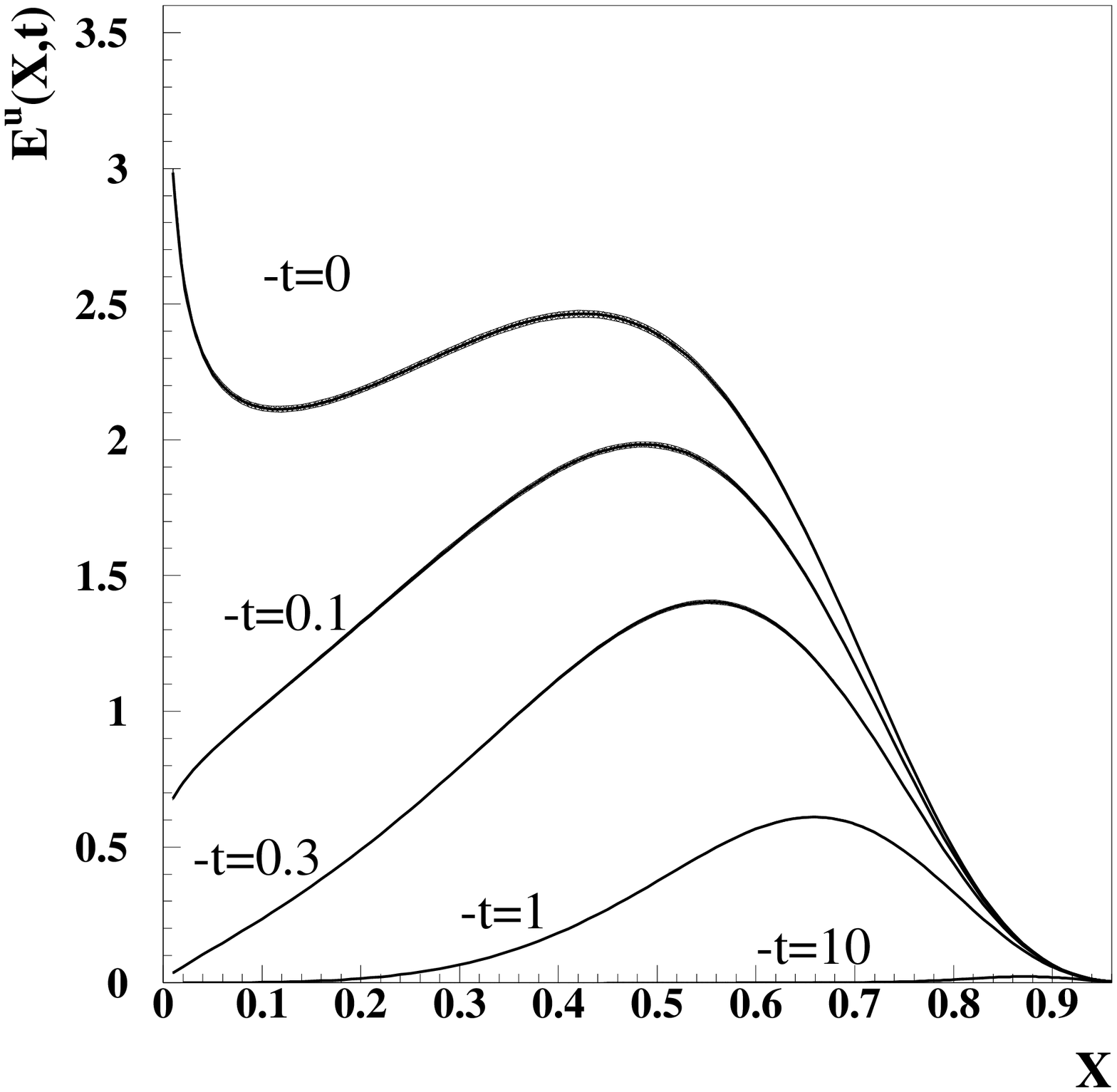}
\includegraphics[width=8.cm]{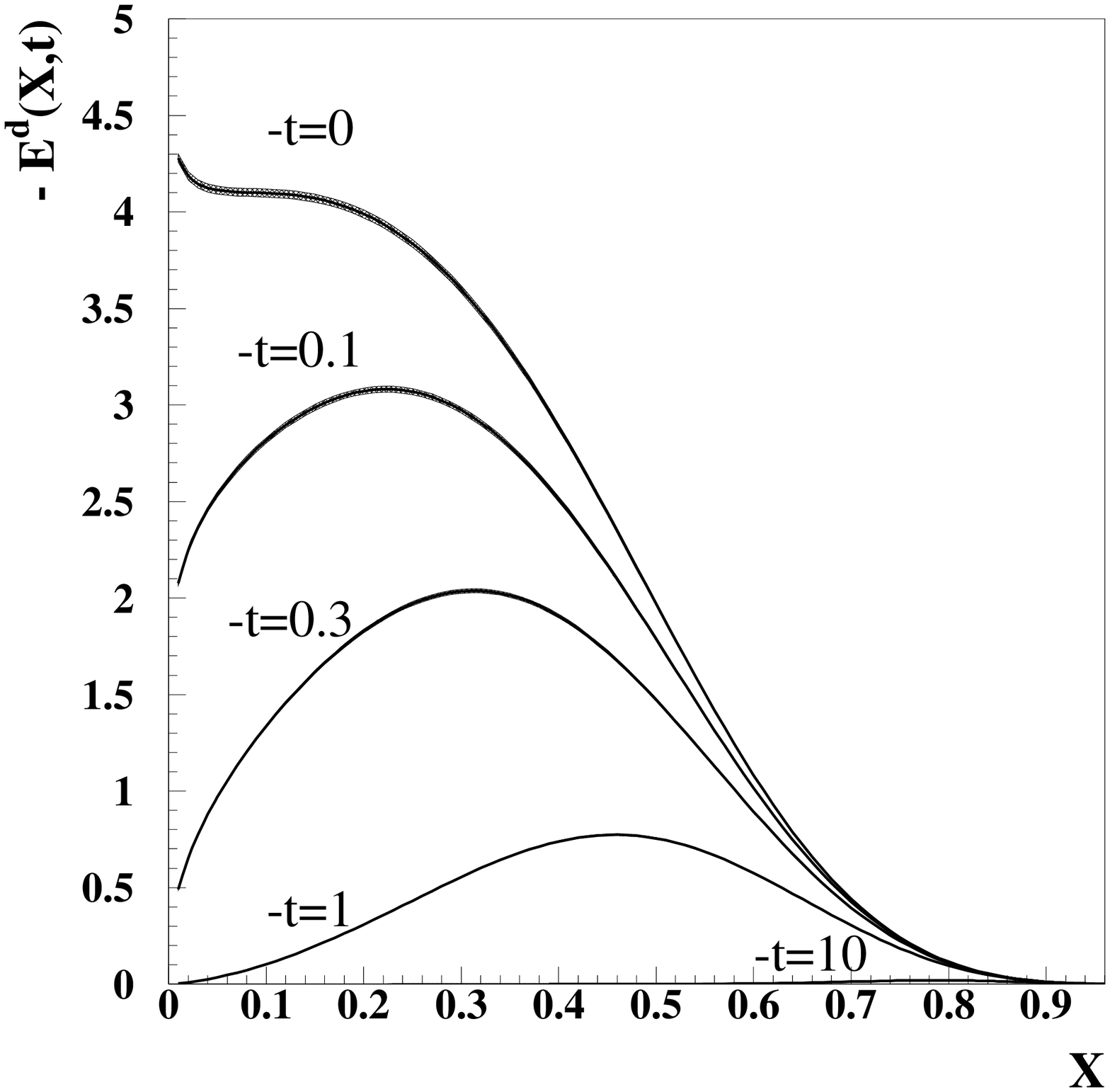}

\caption{The generalized parton distributions, $E_u$ (left panel) and $-E_d$ (right panel) 
obtained from Eq.~(\ref{param1_E}) -- Parametrization I -- 
plotted vs. $X$ at $-t=0,0.1,0.3,1, 10$ GeV$^2$, respectively, at the initial scale
$Q_0^2 = 0.09$ GeV$^2$.}  
\label{fig8}
\end{figure}
%%%%%%%%%%%%%%%%%%%%%%%%%%%%%%%%%%%%%%%%%%%%%%%%%%%%%%%%%%%%%%%%%%%%%%%%%%%%%%%%%%%%%%%%%%%%%% 
%%%%%%%%%%%%%%%%%%%%%%%%%%%%%%%%%%%%%%%%%%%%%%%%%%%%%%%%%%%%%%%%%%%%%%%%%%%%%%%%%%%%%%%%%%%%%% 
\newpage
%%%
%%% FIGURE 9
%%%
\begin{figure}
%{8.5cm}
\includegraphics[width=8.cm]{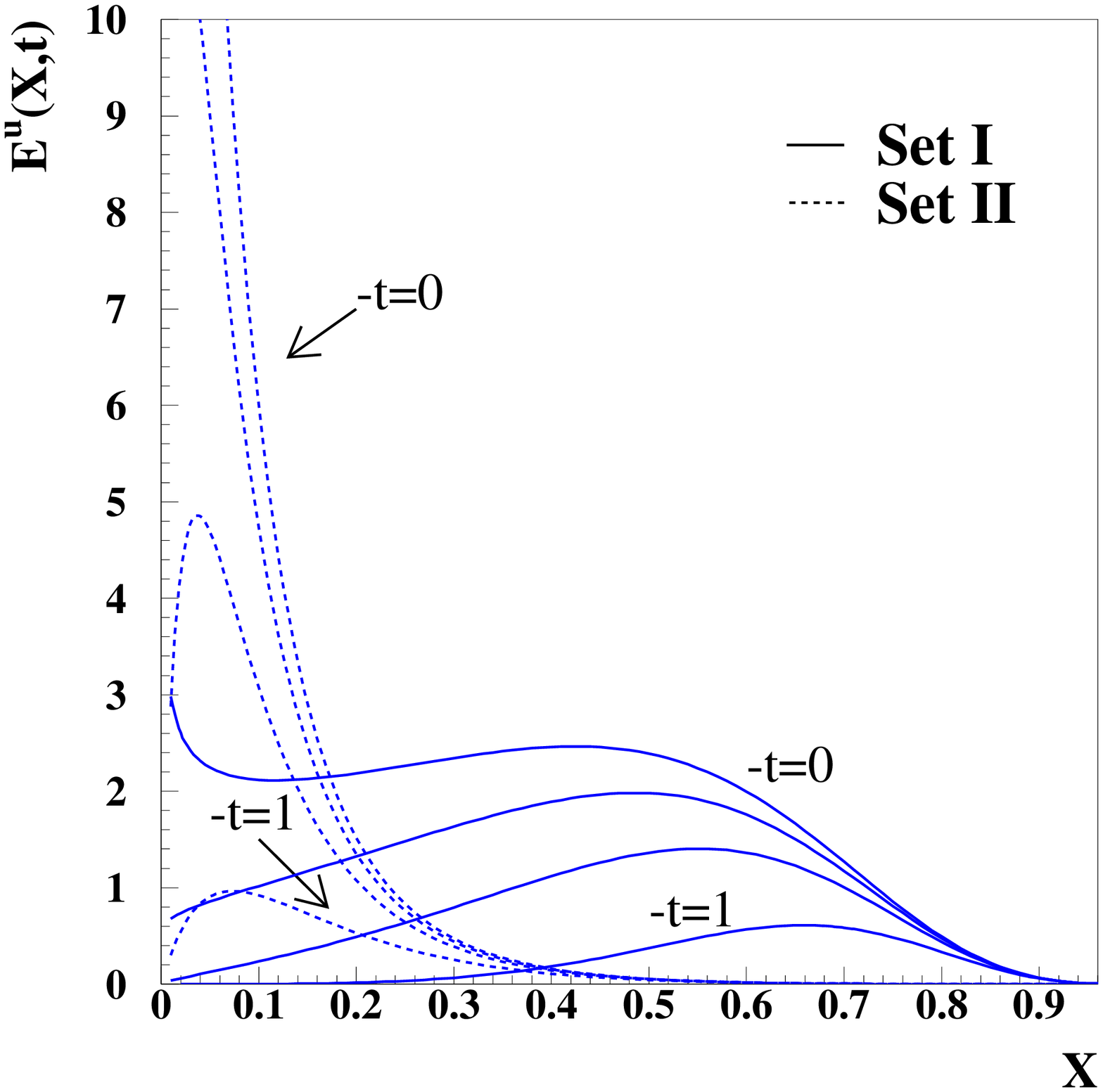}
\includegraphics[width=8.cm]{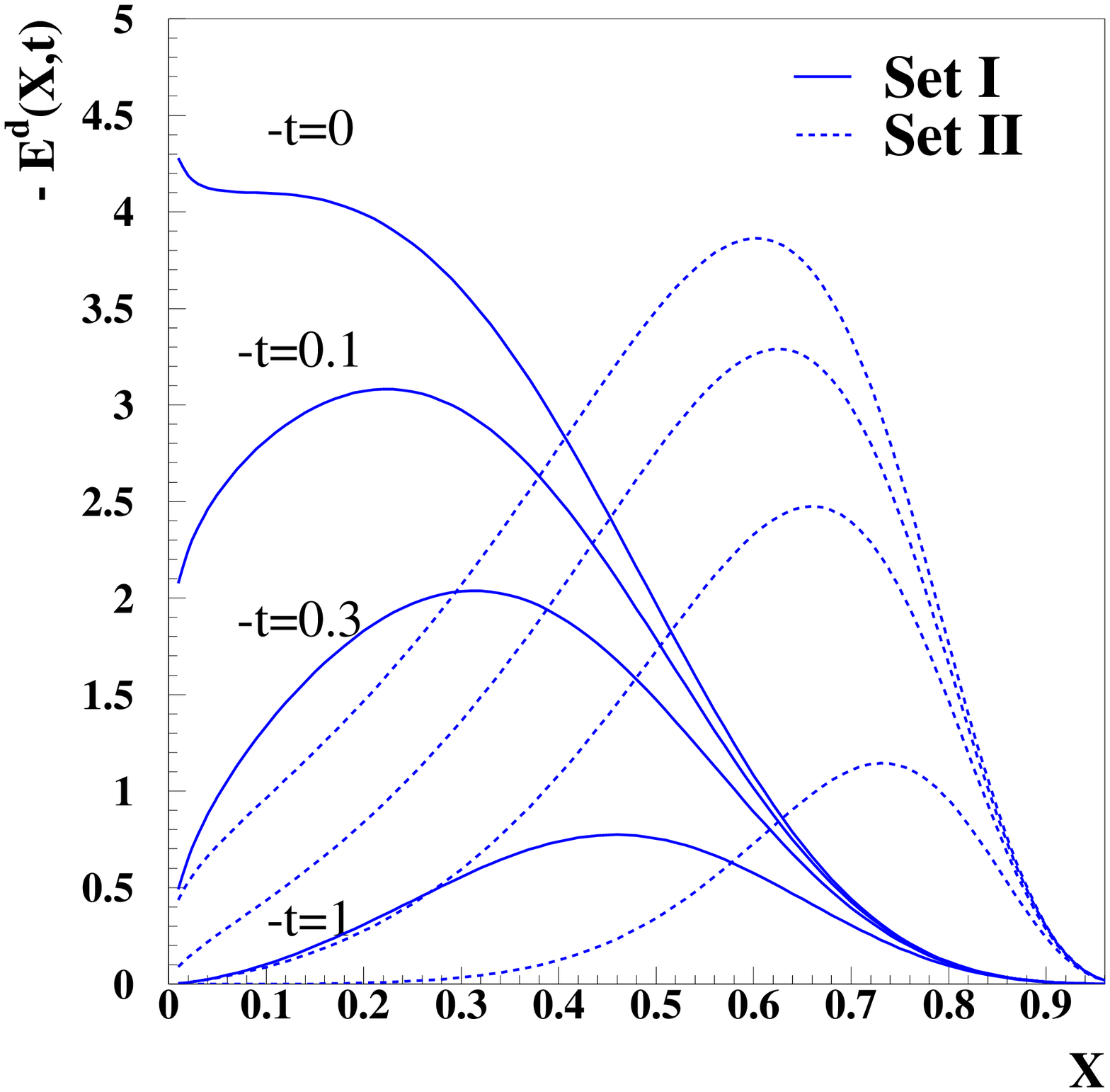}

\caption{Left panel: $E_u$ obtained with Parametrization I (full lines) and II (dashed lines),
respectively, for different values of $-t$: $-t=0,0.1,0.3,1$ GeV$^2$, at the initial scale
$Q_0^2 = 0.09$ GeV$^2$. 
Right panel: the same for $-E_d$ .}
\label{fig9}
\end{figure}
%% 
%%%%%%%%%%%%%%%%%%%%%%%%%%%%%%%%%%%%%%%%%%%%%%%%%%%%%%%%%%%%%%%%%%%%%%%%%%%%%%%%%%%%%%%%%%%%%% 
%%%%%%%%%%%%%%%%%%%%%%%%%%%%%%%%%%%%%%%%%%%%%%%%%%%%%%%%%%%%%%%%%%%%%%%%%%%%%%%%%%%%%%%%%%%%%% 
\newpage
%%%
%%% FIGURE 10
%%%
\begin{figure}
%{8.5cm}
\includegraphics[width=10.cm]{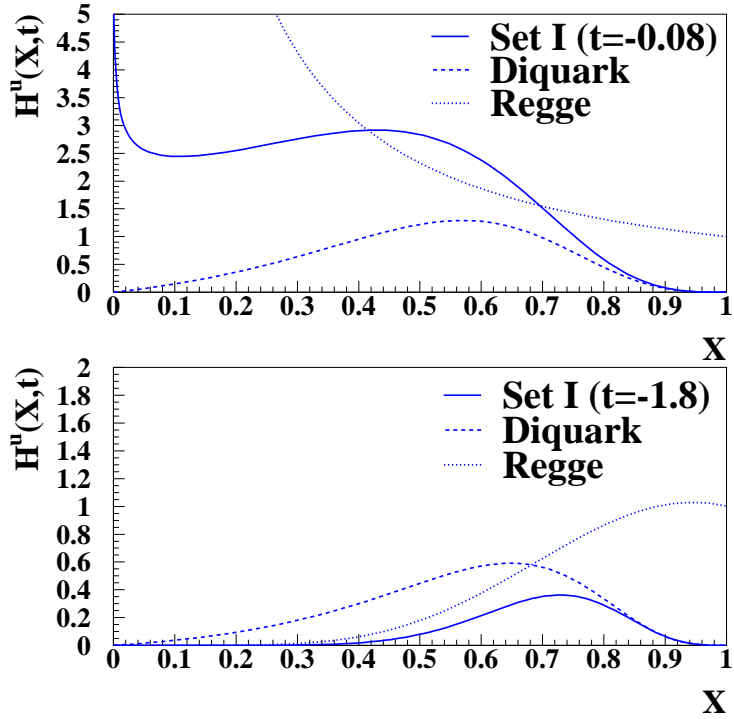}
\caption{Contributions to $H_u$ 
of the diquark term:  $G_{M_X}^{\lambda}$, Eq.~(\ref{gkaava}),
and of the Regge terms: $R =  X^{-\alpha - \beta (1-X)^{p_1} t}$, respectively, at 
$t=-0.08$ GeV$^2$ (top) and $t=-1.8$ GeV$^2$ (bottom). 
From the figure one can see that the form of the GPDs is determined by 
the diquark shape, with an ``envelope'' provided by the Regge term.}
\label{fig10}
\end{figure}
%% 
%%%%%%%%%%%%%%%%%%%%%%%%%%%%%%%%%%%%%%%%%%%%%%%%%%%%%%%%%%%%%%%%%%%%%%%%%%%%%%%%%%%%%%%%%%%%%% 
%%%%%%%%%%%%%%%%%%%%%%%%%%%%%%%%%%%%%%%%%%%%%%%%%%%%%%%%%%%%%%%%%%%%%%%%%%%%%%%%%%%%%%%%%%%%%% 
%%%%%%%%%%%%%%%%%%%%%%%%%%%%%%%%%%%%%%%%%%%%%%%%%%%%%%%%%%%%%%%%%%%%%%%%%%%%%%%%%%%%%%%%%%%%%% 
%%%%%%%%%%%%%%%%%%%%%%%%%%%%%%%%%%%%%%%%%%%%%%%%%%%%%%%%%%%%%%%%%%%%%%%%%%%%%%%%%%%%%%%%%%%%%% 
\newpage
%%%
%%% FIGURE 11
%%%
\begin{figure}
%{8.5cm}
\includegraphics[width=9.cm]{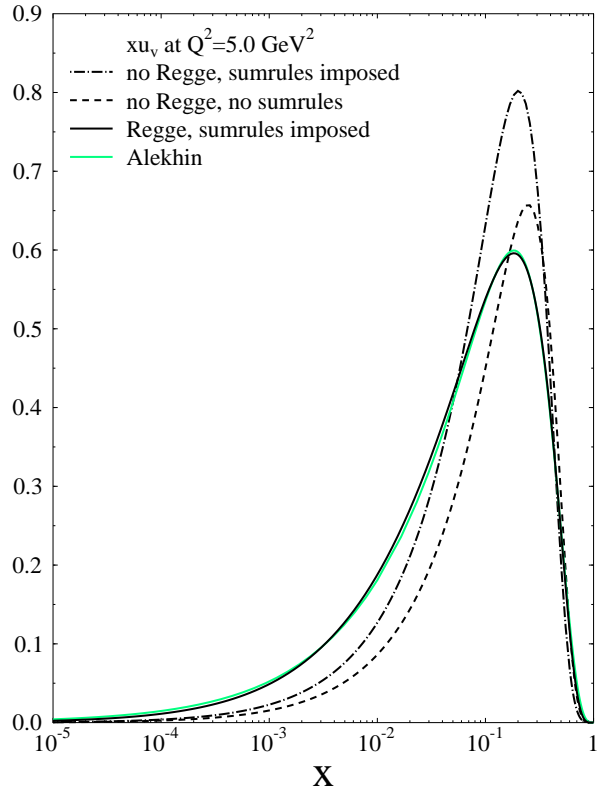}
\caption{(color online) Role of the Regge-motivated term, $R = X^{-\alpha - \beta (1-X)^{p_1} t}$, 
in $H_u$, in order to accomplish a quantitative fit of the PDF $u_v(X,Q^2)$.} 
\label{fig11}
\end{figure}
%% 
%%%%%%%%%%%%%%%%%%%%%%%%%%%%%%%%%%%%%%%%%%%%%%%%%%%%%%%%%%%%%%%%%%%%%%%%%%%%%%%%%%%%%%%%%%%%%% 
%%%%%%%%%%%%%%%%%%%%%%%%%%%%%%%%%%%%%%%%%%%%%%%%%%%%%%%%%%%%%%%%%%%%%%%%%%%%%%%%%%%%%%%%%%%%%% 
\newpage
%%%
%%% FIGURE 12
%%%
\begin{figure}
%{8.5cm}
\includegraphics[width=10.cm]{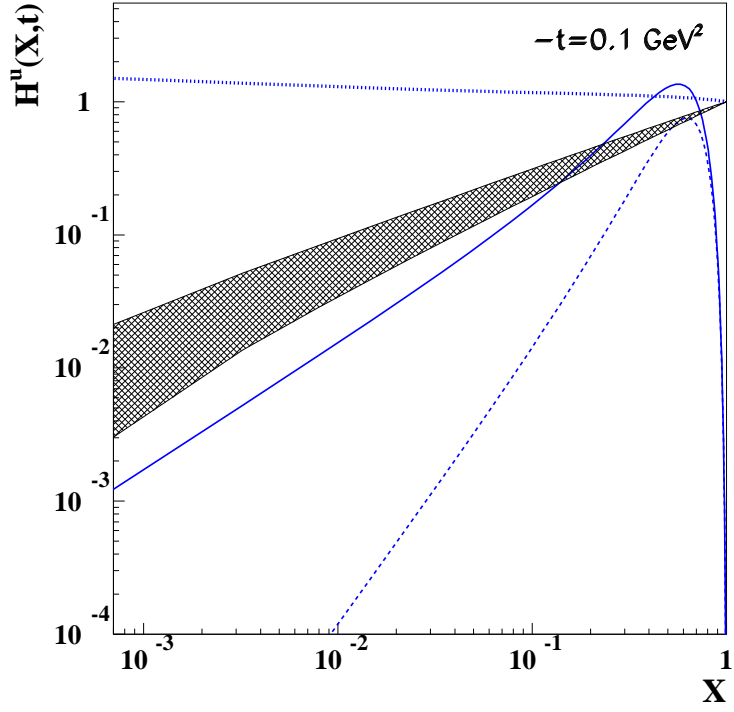}
\caption{(color online) Role of Regge-motivated term in 
Eqs.~(\protect\ref{param1_H},\protect\ref{param2_H}). Thick dotted line: $X^{-\alpha-\beta(1-x)^{p_1} t}$, 
with fit parameters from Tables II and III; dashed line: quark-diquark term, $G_{M_X}^\lambda$;
Full line: total value, Eqs.~(\protect\ref{param1_H},\protect\ref{param2_H}). Shaded area: 
Regge term: $X^{-\alpha^\prime-\beta^\prime t}$, with fit parameters from phenomenological
Regge fits.   
All curves were calculated at $t=-0.1$ GeV$^2$.}
\label{fig12}
\end{figure}
%%%%%%%%%%%%%%%%%%%%%%%%%%%%%%%%%%%%%%%%%%%%%%%%%%%%%%%%%%%%%%%%%%%%%%%%%%%%%%%%%%%%%%%%%%%%%% 
%%%%%%%%%%%%%%%%%%%%%%%%%%%%%%%%%%%%%%%%%%%%%%%%%%%%%%%%%%%%%%%%%%%%%%%%%%%%%%%%%%%%%%%%%%%%%% 
%%%%%%%%%%%%%%%%%%%%%%%%%%%%%%%%%%%%%%%%%%%%%%%%%%%%%%%%%%%%%%%%%%%%%%%%%%%%%%%%%%%%%%%%%%%%%% 
%%%%%%%%%%%%%%%%%%%%%%%%%%%%%%%%%%%%%%%%%%%%%%%%%%%%%%%%%%%%%%%%%%%%%%%%%%%%%%%%%%%%%%%%%%%%%% 
%%%%%%%%%%%%%%%%%%%%%%%%%%%%%%%%%%%%%%%%%%%%%%%%%%%%%%%%%%%%%%%%%%%%%%%%%%%%%%%%%%%%%%%%%%%%%% 
%%%%%%%%%%%%%%%%%%%%%%%%%%%%%%%%%%%%%%%%%%%%%%%%%%%%%%%%%%%%%%%%%%%%%%%%%%%%%%%%%%%%%%%%%%%%%% 
\newpage
%%%
%%% FIGURE 13
%%%
\begin{figure}
%{8.5cm}
\includegraphics[width=10.cm]{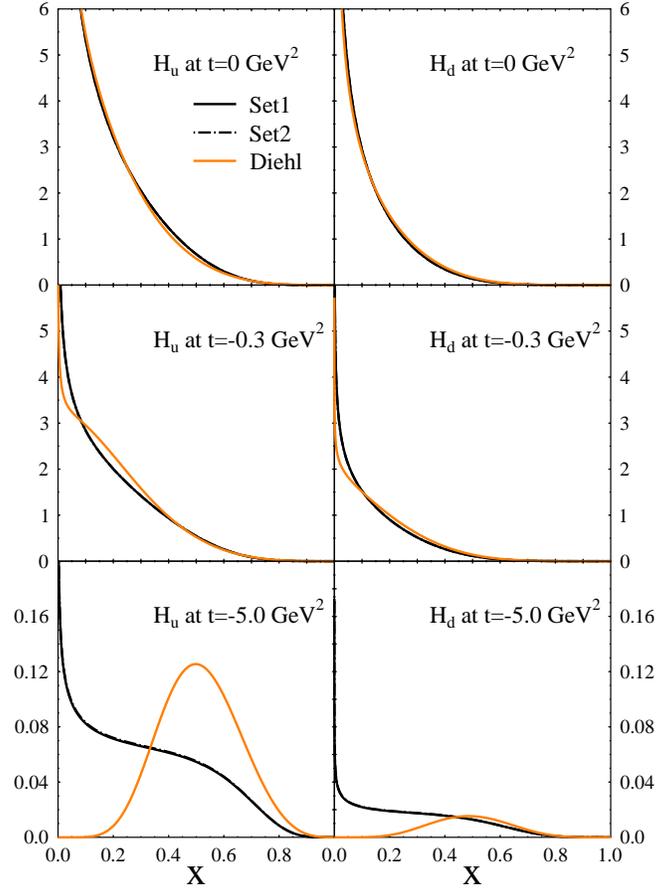}
\caption{(color online) Comparison with the quantitative extraction of GPDs from data from Ref.~\cite{DieKro1}.
$H_u$ (left) and $H_d$ (right) 
as a function of $X$ for $-t=0, 0.3, 5$ GeV$^2$, for our Parametrizations I and II,
respectively evolved at $Q^2=4$ GeV$^2$ used in  Ref.~\cite{DieKro1} While a clear agreement
is seen at low values of $t$, this becomes worse at larger values of $t$.}
\label{fig13}
\end{figure}
%%%%%%%%%%%%%%%%%%%%%%%%%%%%%%%%%%%%%%%%%%%%%%%%%%%%%%%%%%%%%%%%%%%%%%%%%%%%%%%%%%%%%%%%%%%%%% 
%%%%%%%%%%%%%%%%%%%%%%%%%%%%%%%%%%%%%%%%%%%%%%%%%%%%%%%%%%%%%%%%%%%%%%%%%%%%%%%%%%%%%%%%%%%%%% 
\newpage
%%%
%%% FIGURE 14
%%%
\begin{figure}
%{8.5cm}
\includegraphics[width=10.cm]{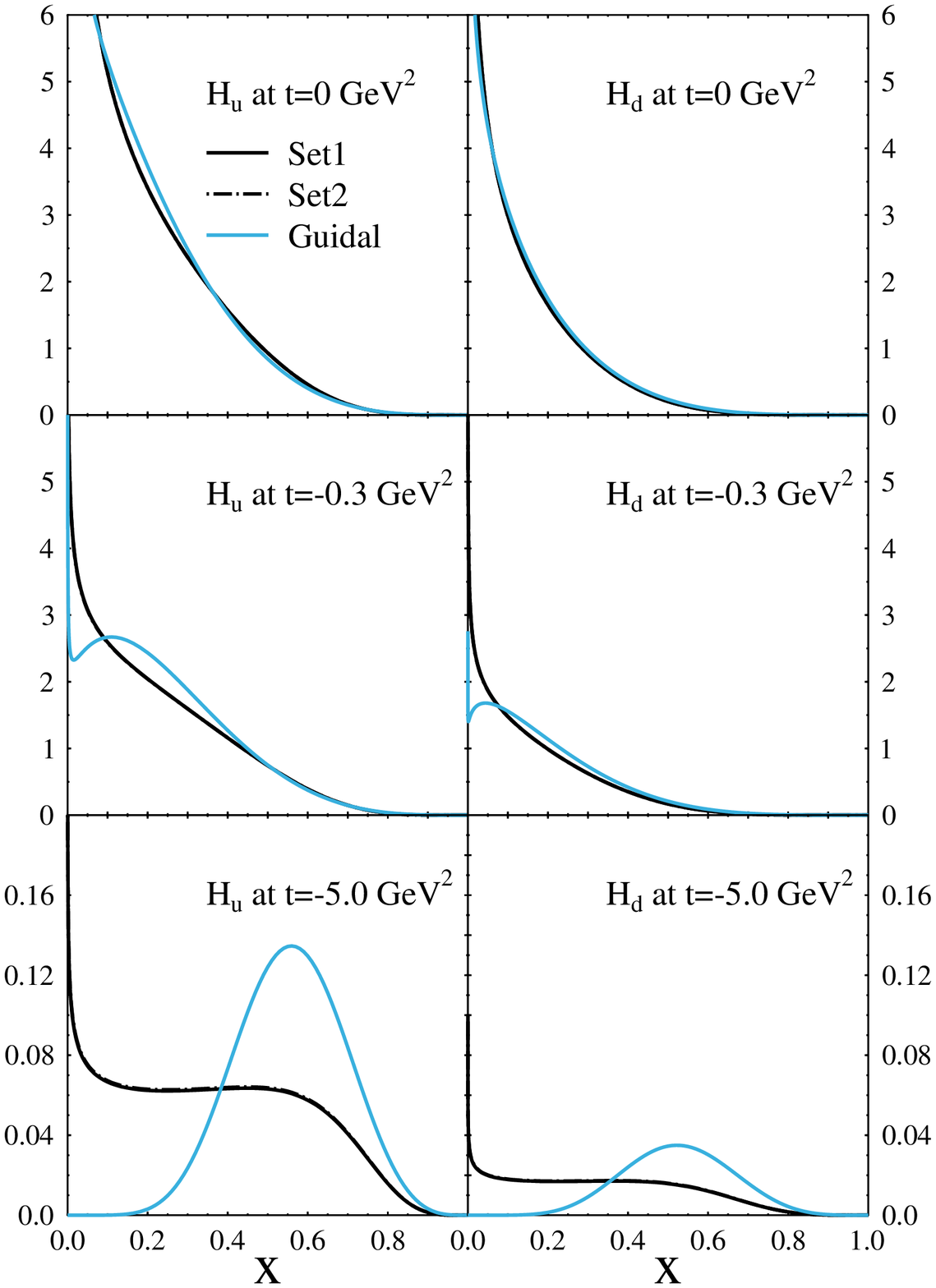}
\caption{(color online) 
Comparison with the quantitative extraction of GPDs from data from Ref.~\cite{VandH1} at 
$Q^2 = 1$ GeV$^2$. Notations as in Fig.~\ref{fig13}.}
\label{fig14}
\end{figure}
%%%%%%%%%%%%%%%%%%%%%%%%%%%%%%%%%%%%%%%%%%%%%%%%%%%%%%%%%%%%%%%%%%%%%%%%%%%%%%%%%%%%%%%%%%%%%% 
%%%%%%%%%%%%%%%%%%%%%%%%%%%%%%%%%%%%%%%%%%%%%%%%%%%%%%%%%%%%%%%%%%%%%%%%%%%%%%%%%%%%%%%%%%%%%% 
%%%%%%%%%%%%%%%%%%%%%%%%%%%%%%%%%%%%%%%%%%%%%%%%%%%%%%%%%%%%%%%%%%%%%%%%%%%%%%%%%%%%%%%%%%%%%% 
%%%%%%%%%%%%%%%%%%%%%%%%%%%%%%%%%%%%%%%%%%%%%%%%%%%%%%%%%%%%%%%%%%%%%%%%%%%%%%%%%%%%%%%%%%%%%% 
\newpage
%%%
%%% FIGURE 15
%%%
\begin{figure}
%{8.5cm}
\includegraphics[width=10.cm]{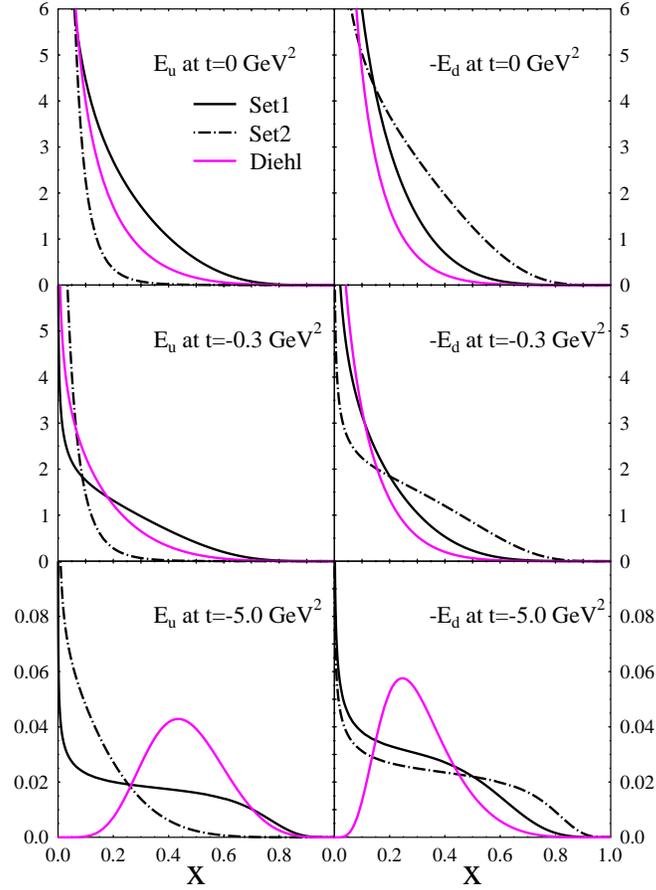}
\caption{(color online) Comparison with the quantitative extraction of GPDs from data from 
Ref.~\cite{DieKro1}. $E_u$ (left) and $-E_d$ (right) 
as a function of $X$ for $-t=0, 0.3, 5$ GeV$^2$, for our Parametrizations I and II,
respectively evolved at $Q^2=4$ GeV$^2$ used in  Ref.~\cite{DieKro1}.}
\label{fig15}
\end{figure}
\clearpage
%%%%%%%%%%%%%%%%%%%%%%%%%%%%%%%%%%%%%%%%%%%%%%%%%%%%%%%%%%%%%%%%%%%%%%%%%%%%%%%%%%%%%%%%%%%%%% 
%%%%%%%%%%%%%%%%%%%%%%%%%%%%%%%%%%%%%%%%%%%%%%%%%%%%%%%%%%%%%%%%%%%%%%%%%%%%%%%%%%%%%%%%%%%%%% 
%%%%%%%%%%%%%%%%%%%%%%%%%%%%%%%%%%%%%%%%%%%%%%%%%%%%%%%%%%%%%%%%%%%%%%%%%%%%%%%%%%%%%%%%%%%%%% 
%%%%%%%%%%%%%%%%%%%%%%%%%%%%%%%%%%%%%%%%%%%%%%%%%%%%%%%%%%%%%%%%%%%%%%%%%%%%%%%%%%%%%%%%%%%%%% 
\newpage
%%%
%%% FIGURE 16
%%%
\begin{figure}
%{8.5cm}
\includegraphics[width=10.cm]{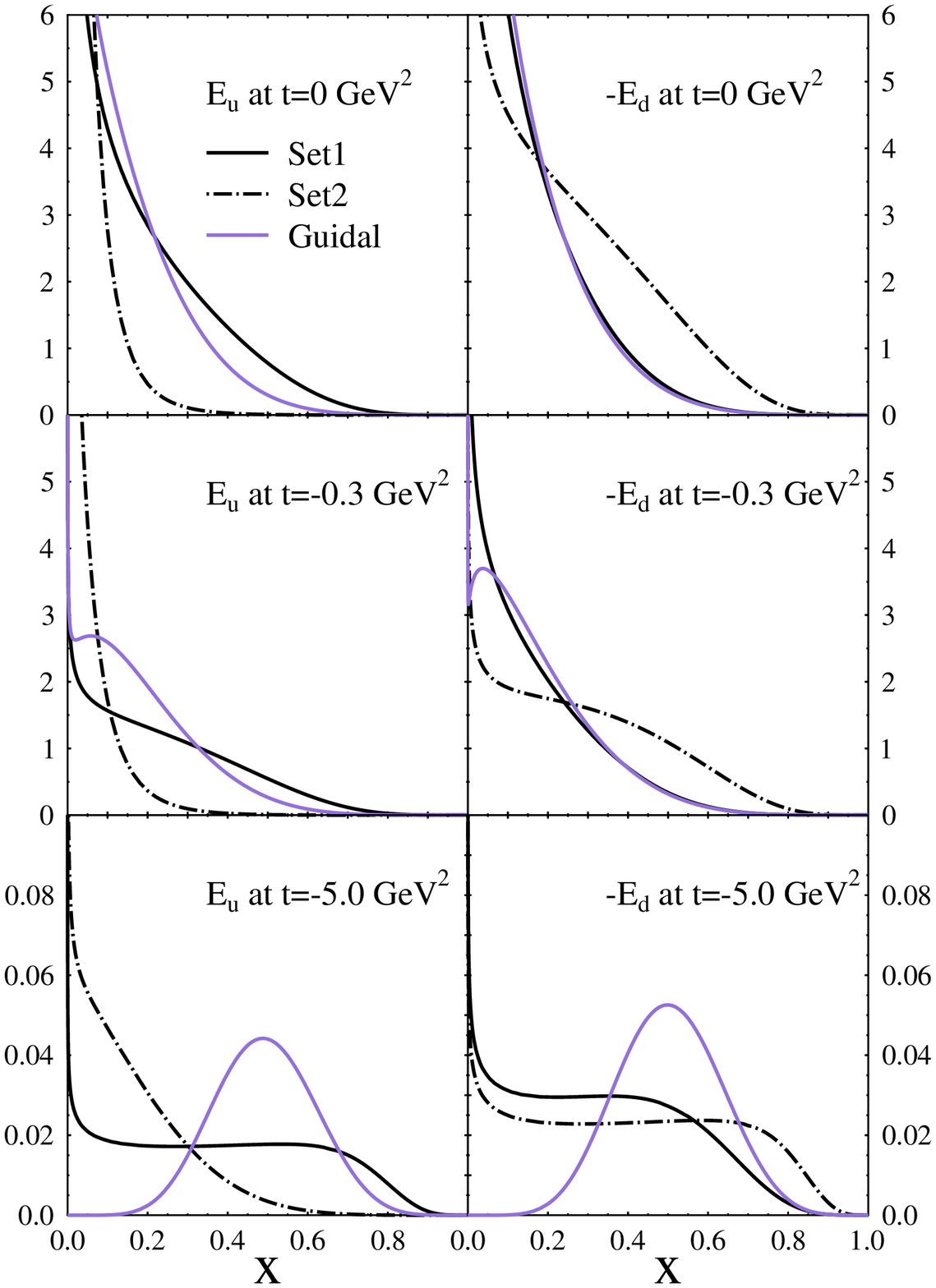}
\caption{(color online) Comparison with the quantitative extraction of GPDs from data from Ref.~\cite{VandH1}
at $Q^2 = 1$ GeV$^2$. Notations as in Fig.~\ref{fig15}.}
\label{fig16}
\end{figure}
%%%%%%%%%%%%%%%%%%%%%%%%%%%%%%%%%%%%%%%%%%%%%%%%%%%%%%%%%%%%%%%%%%%%%%%%%%%%%%%%%%%%%%%%%%%%%% 
%%%%%%%%%%%%%%%%%%%%%%%%%%%%%%%%%%%%%%%%%%%%%%%%%%%%%%%%%%%%%%%%%%%%%%%%%%%%%%%%%%%%%%%%%%%%%% 
%%%%%%%%%%%%%%%%%%%%%%%%%%%%%%%%%%%%%%%%%%%%%%%%%%%%%%%%%%%%%%%%%%%%%%%%%%%%%%%%%%%%%%%%%%%%%% 
%%%%%%%%%%%%%%%%%%%%%%%%%%%%%%%%%%%%%%%%%%%%%%%%%%%%%%%%%%%%%%%%%%%%%%%%%%%%%%%%%%%%%%%%%%%%%% 
\newpage
%%%
%%% FIGURE 17
%%%
\begin{figure}
%{8.5cm}
\includegraphics[width=8.cm]{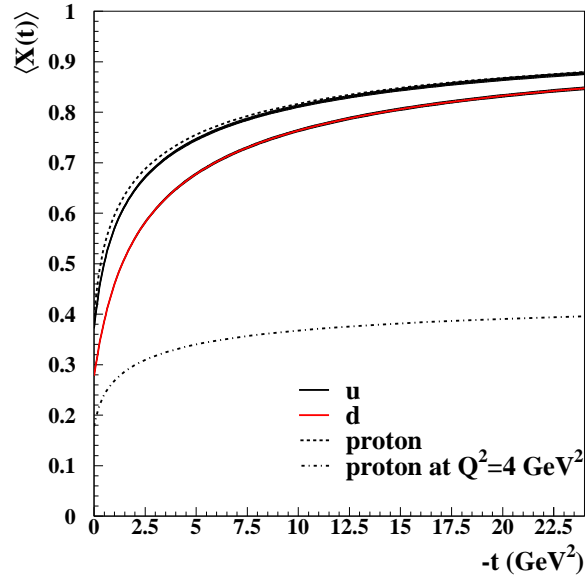}
\caption{(color online) Average value of $X$, Eq.~(\ref{xave}, plotted vs. $-t$ at $Q^2=Q^2_0$ for the 
$u$ and $d$ quarks contributions to the proton Dirac form factor, Eq.~(\ref{FF2}). The average vale 
of $X$ for the proton is also shown at $Q^2=4$ GeV$^2$ (dot-dashed line).}  
\label{fig17}
\end{figure}
%%%%%%%%%%%%%%%%%%%%%%%%%%%%%%%%%%%%%%%%%%%%%%%%%%%%%%%%%%%%%%%%%%%%%%%%%%%%%%%%%%%%%%%%%%%%%% 
%%%%%%%%%%%%%%%%%%%%%%%%%%%%%%%%%%%%%%%%%%%%%%%%%%%%%%%%%%%%%%%%%%%%%%%%%%%%%%%%%%%%%%%%%%%%%%%%%%%%%%%%%%%%%%%%% 
%%%%%%%%%%%%%%%%%%%%%%%%%%%%%%%%%%%%%%%%%%%%%%%%%%%%%%%%%%%%%%%%%%%%%%%%%%%%%%%%%%%%%%%%%%%%%% 
%%%%%%%%%%%%%%%%%%%%%%%%%%%%%%%%%%%%%%%%%%%%%%%%%%%%%%%%%%%%%%%%%%%%%%%%%%%%%%%%%%%%%%%%%%%%%% 
%%%%%%%%%%%%%%%%%%%%%%%%%%%%%%%%%%%%%%%%%%%%%%%%%%%%%%%%%%%%%%%%%%%%%%%%%%%%%%%%%%%%%%%%%%%%%% 
\newpage
%%%
%%% FIGURE 18
%%%
\begin{figure}
%{8.5cm}
\includegraphics[width=8.cm]{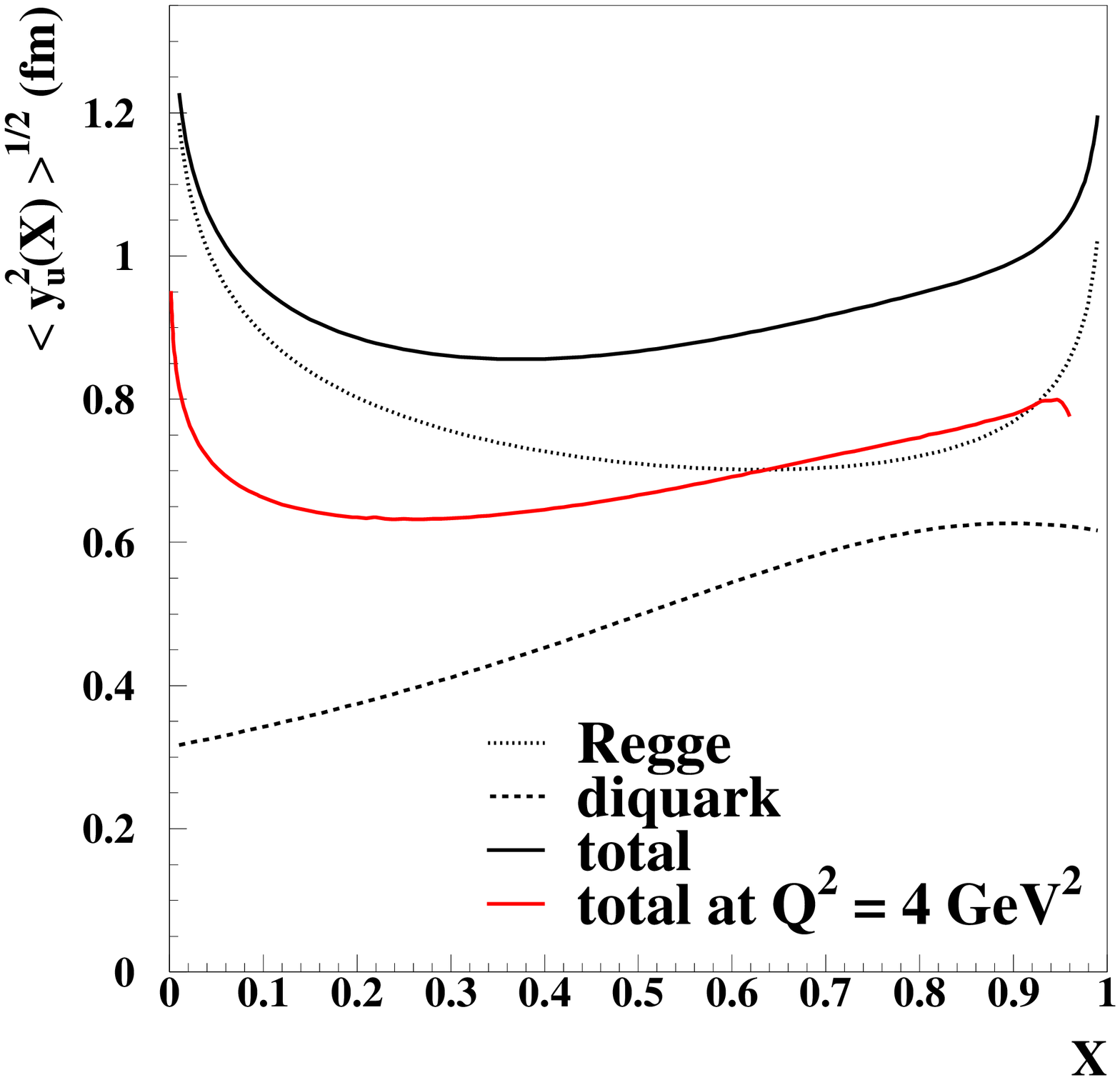}
\includegraphics[width=8.cm]{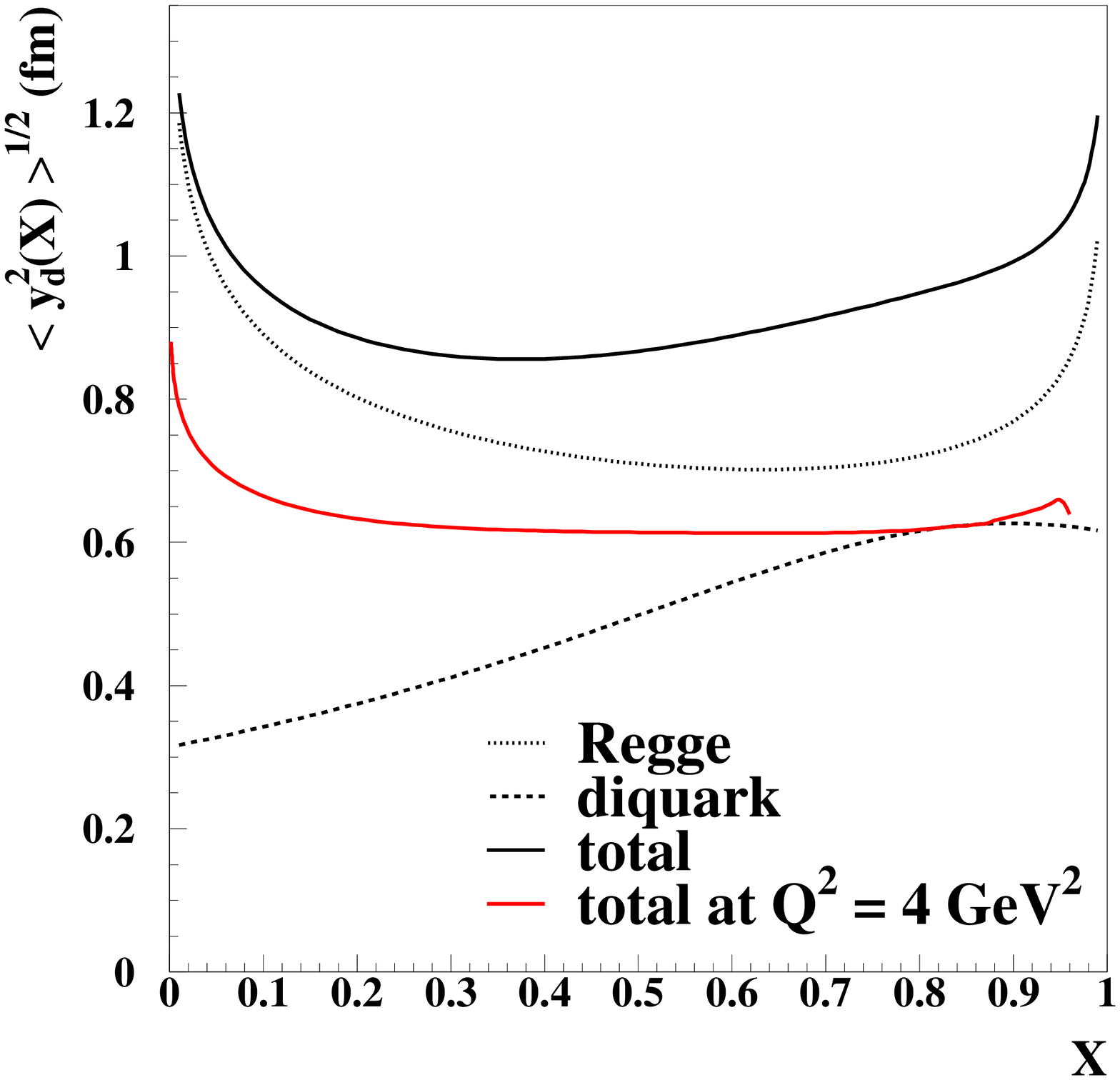}
\caption{(color online) 
$\langle y_u^2(X) \rangle$ (left panel) and $\langle y_d^2(X) \rangle$ (right panel) 
in our model, plotted vs. $X$. The contribution of the Regge and diquark term, respectively 
(see text) are shown separately along with the total result, at the initial scale. 
The total result is then evolved to $Q^2=4$ GeV$^2$.   
}
\label{fig18}
\end{figure}
%%%%%%%%%%%%%%%%%%%%%%%%%%%%%%%%%%%%%%%%%%%%%%%%%%%%%%%%%%%%%%%%%%%%%%%%%%%%%%%%%%%%%%%%%%%%%% 
%%%%%%%%%%%%%%%%%%%%%%%%%%%%%%%%%%%%%%%%%%%%%%%%%%%%%%%%%%%%%%%%%%%%%%%%%%%%%%%%%%%%%%%%%%%%%% 
%%%%%%%%%%%%%%%%%%%%%%%%%%%%%%%%%%%%%%%%%%%%%%%%%%%%%%%%%%%%%%%%%%%%%%%%%%% 
%%%%%%%%%%%%%%%%%%%%%%%%%%%%%%%%%%%%%%%%%%%%%%%%%%%%%%%%%%%%%%%%%%%%%%%%%%%%%%%%%%%%%%%%%%%%% 
%%%%%%%%%%%%%%%%%%%%%%%%%%%%%%%%%%%%%%%%%%%%%%%%%%%%%%%%%%%%%%%%%%%%%%%%%%%%%%%%%%%%%%%%%%%%%% 
%%%%%%%%%%%%%%%%%%%%%%%%%%%%%%%%%%%%%%%%%%%%%%%%%%%%%%%%%%%%%%%%%%%%%%%%%%%%%%%%%%%%%%%%%%%%%% 
%%%
%%% FIGURE 19
%%%
\begin{figure}
%{8.5cm}
\includegraphics[width=8.cm]{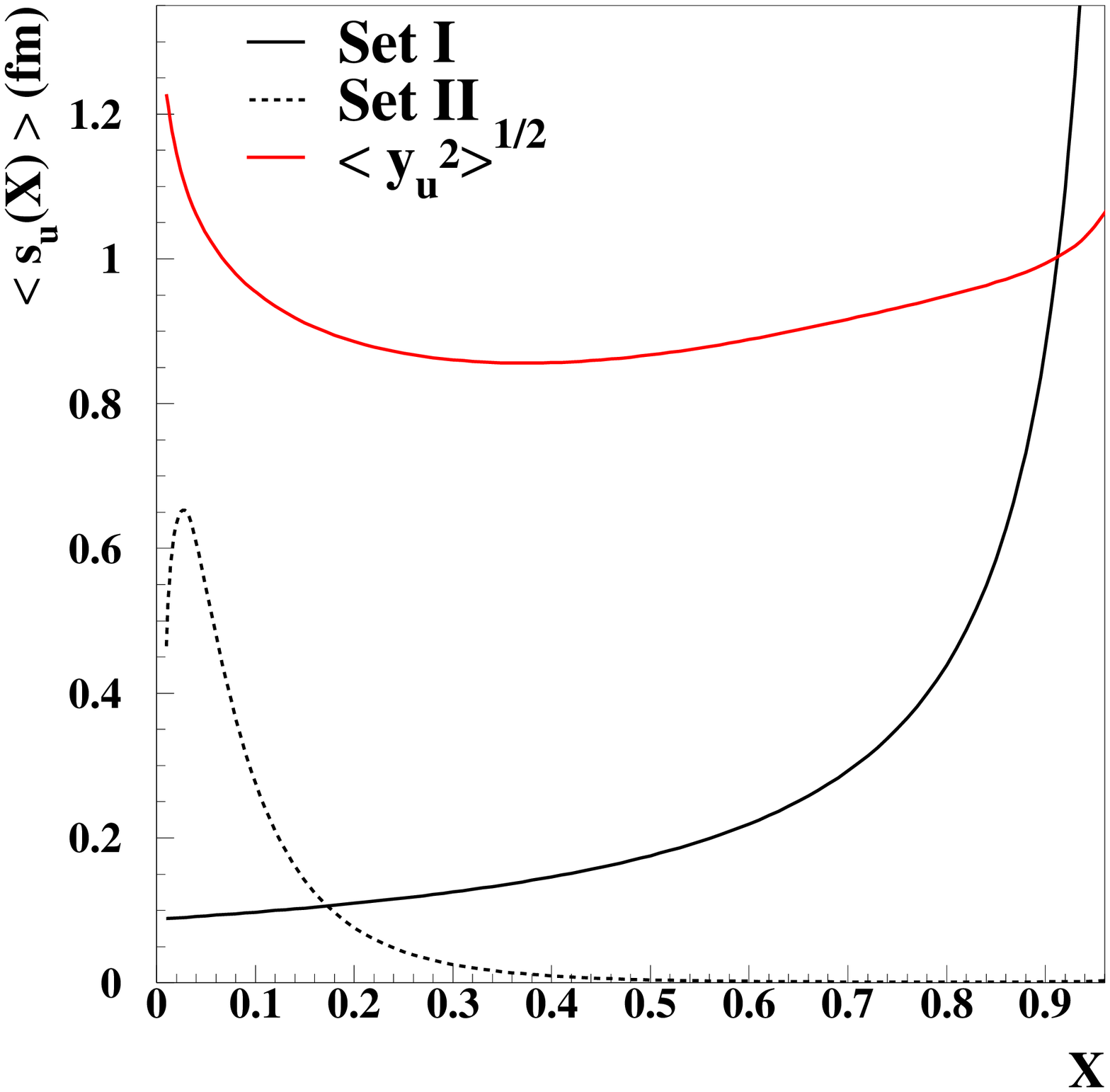}
\includegraphics[width=8.cm]{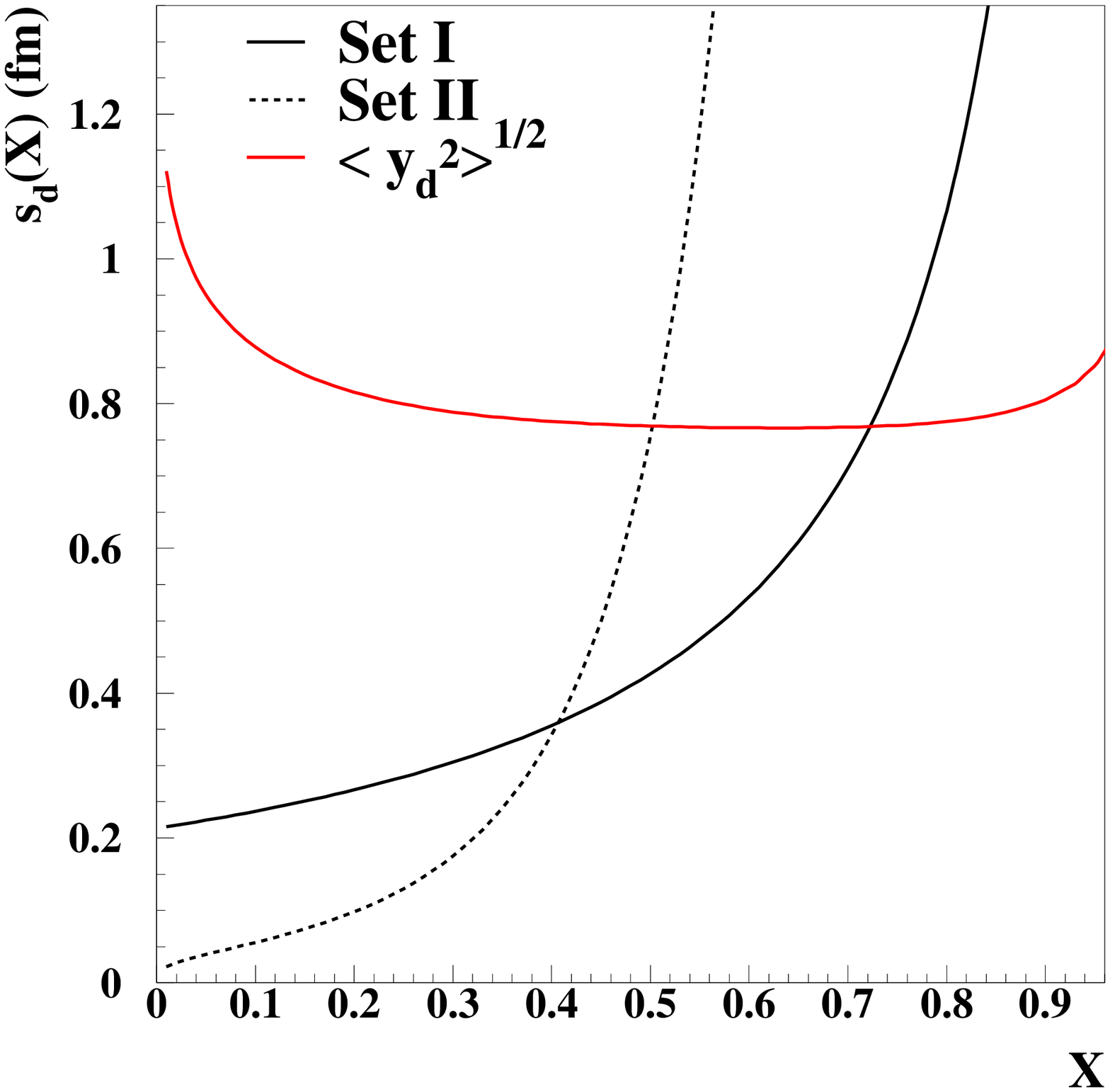}
\caption{(color online) $\langle s_u(X) \rangle$ (left panel) and $\langle s_d(X) \rangle$ 
(right panel) in our model, plotted vs. $X$. All curves are at the initial scale, $Q_0^2$. 
The average interparton distance: $\langle y_q^2(X) \rangle^{1/2}$ 
is shown for comparison.}
\label{fig19}
\end{figure}

\newpage
%%%%%%%%%%%%%%%%%%%%%%%%%%%%%%%%%%%%%%%%%%%%%%%%%%%%%%%%%%%%%%%%%%%%%%%%%%%%%%%%%%%%%%%%%%%%%% 
%%%%%%%%%%%%%%%%%%%%%%%%%%%%%%%%%%%%%%%%%%%%%%%%%%%%%%%%%%%%%%%%%%%%%%%%%%%%%%%%%%%%%%%%%%%%%% 
%%%
%%% FIGURE 20
%%%
\begin{figure}
%{8.5cm}
\includegraphics[width=8.cm]{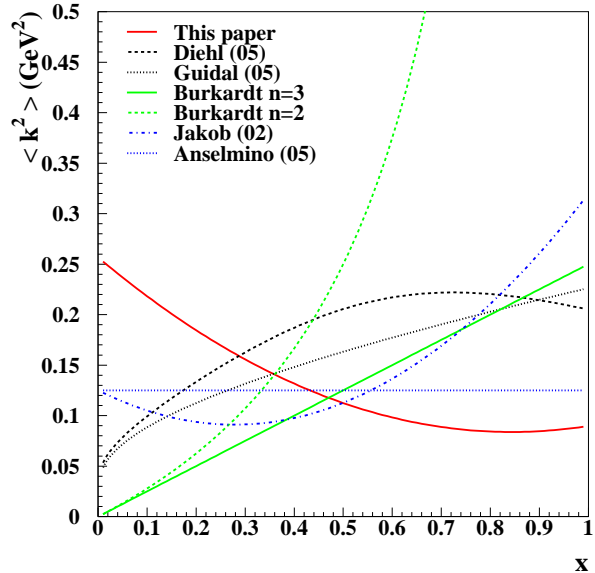}
\caption{(color online) The average intrinsic transverse momentum, Eq.~(\ref{kperp}), compared
to the value extracted from GPD parametrizations (Refs.~\cite{DieKro1}, \cite{VandH1}); to the 
$t$-dependence conjecture of Ref.~\cite{Burk2}, and to models used in Semi-Inclusive DIS (SIDIS),
Refs.~\cite{Jakob:1997wg} and \cite{anselmino}.} 
\label{fig20}
\end{figure}
%%%%%%%%%%%%%%%%%%%%%%%%%%%%%%%%%%%%%%%%%%%%%%%%%%%%%%%%%%%%%%%%%%%%%%%%%%%%%%%%%%%%%%%%%%%%%% 
%%%%%%%%%%%%%%%%%%%%%%%%%%%%%%%%%%%%%%%%%%%%%%%%%%%%%%%%%%%%%%%%%%%%%%%%%%%%%%%%%%%%%%%%%%%%%% 
%%%%%%%%%%%%%%%%%%%%%%%%%%%%%%%%%%%%%%%%%%%%%%%%%%%%%%%%%%%%%%%%%%%%%%%%%%%%%%%%%%%%%%%%%%%%%% 
%%%
%%% FIGURE 21
%%%
\begin{figure}
%{8.5cm}
\includegraphics[width=10.cm]{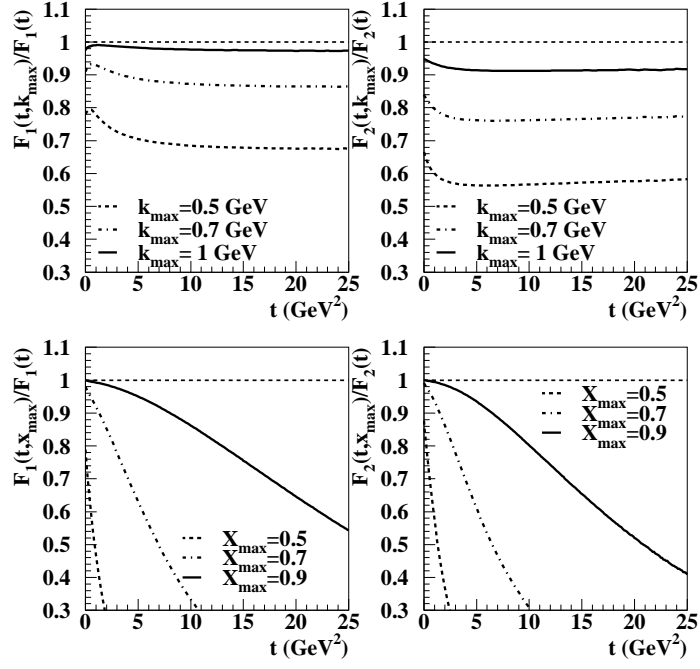}
\caption{Contribution of both the intrinsic transverse momentum components (upper panels)
and $X$ components (lower panels) 
to the proton form factors, $F_1^p(t)$ (left), and $F_2^p(t)$ (right).}
\label{fig21}
\end{figure}
%%%%%%%%%%%%%%%%%%%%%%%%%%%%%%%%%%%%%%%%%%%%%%%%%%%%%%%%%%%%%%%%%%%%%%%%%%%%%%%%%%%%%%%%%%%%%%
%%%%%%%%%%%%%%%%%%%%%%%%%%%%%%%%%%%%%%%%%%%%%%%%%%%%%%%%%%%%%%%%%%%%%%%%%%%%%%%%%%%%%%%%%%%%%% 
%%%%%%%%%%%%%%%%%%%%%%%%%%%%%%%%%%%%%%%%%%%%%%%%%%%%%%%%%%%%%%%%%%%%%%%%%%%%%%%%%%%%%%%%%%%%%% 
%%%
%%% FIGURE 22
%%%
\begin{figure}
%{8.5cm}
\includegraphics[width=10.cm]{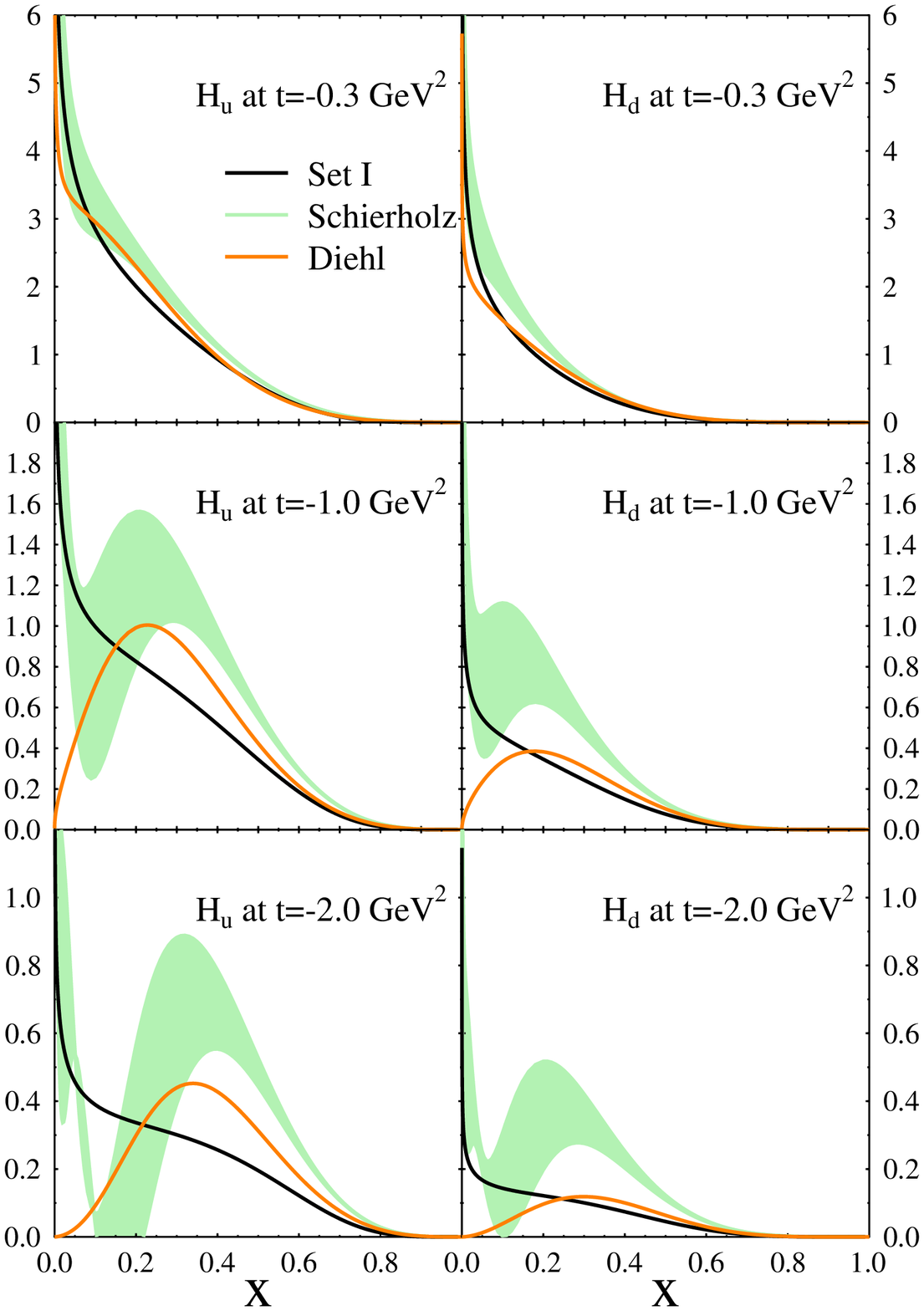}
\caption{(color online) Comparison of both the results of our analysis, and of 
Ref.~\cite{DieKro1}, with an extraction 
of GPDs from lattice calculations according to the prescription of Ref.~\cite{Schierholz}.
The band includes an estimate of the error from lattice calculations.}
\label{fig22}
\end{figure}
%%%%%%%%%%%%%%%%%%%%%%%%%%%%%%%%%%%%%%%%%%%%%%%%%%%%%%%%%%%%%%%%%%%%%%%%%%%%%%%%%%%%%%%%%%%%%% 


\begin{thebibliography}{90}
%%%%% INTRO	
\bibitem{DMul1} D.~Muller, D.~Robaschik, B.~Geyer, F.~M.~Dittes and J.~Horejsi,
Fortsch.\ Phys.\  {\bf 42}, 101 (1994)

\bibitem{Ji1} X.~D.~Ji, Phys.\ Rev.\ D {\bf 55}, 7114 (1997)

\bibitem{Rad1} A.~V.~Radyushkin,
Phys.\ Rev.\ D {\bf 56}, 5524 (1997)

\bibitem{Bur} M.~Burkardt,
Int.\ J.\ Mod.\ Phys.\ A {\bf 18}, 173 (2003); {\it ibid}
Phys.\ Rev.\ D {\bf 62}, 071503 (2000)
[Erratum-ibid.\ D {\bf 66}, 119903 (2002)].

\bibitem{PDG} W.-M. Yao et al., J. Phys. G 33, 1 (2006), ``2006 Review of Particle Physics''.

\bibitem{BelDVCS} A.~V.~Belitsky, D.~Mueller and A.~Kirchner,
  Nucl.\ Phys.\ B {\bf 629}, 323 (2002)
  
\bibitem{Die_rev} M.~Diehl, Phys.\ Rept.\  {\bf 388}, 41 (2003).

\bibitem{BelRad} A.~V.~Belitsky and A.~V.~Radyushkin,
  Phys.\ Rept.\  {\bf 418}, 1 (2005)

\bibitem{Sto} P.~Stoler,
  Phys.\ Rev.\ D {\bf 65}, 053013 (2002)
 
\bibitem{GSL} A.~Gardestig, A.~P.~Szczepaniak and J.~T.~Londergan,
  Phys.\ Rev.\ D {\bf 68}, 034005 (2003)

\bibitem{afa} A.V. Afanasev, hep-ph/9808291; hep-ph/9910565.

\bibitem{LiuTan1} S.~Liuti and S.~K.~Taneja,
  Phys.\ Rev.\ D {\bf 70}, 074019 (2004)

\bibitem{DieKro1} M.~Diehl, T.~Feldmann, R.~Jakob and P.~Kroll,
  Eur.\ Phys.\ J.\ C {\bf 39}, 1 (2005)
 
\bibitem{VandH1} M.~Guidal, M.~V.~Polyakov, A.~V.~Radyushkin and M.~Vanderhaeghen,
  Phys.\ Rev.\ D {\bf 72}, 054013 (2005)

\bibitem{RadDD} A.~V.~Radyushkin,
  arXiv:hep-ph/0101225.
  
\bibitem{VGG} M.~Vanderhaeghen, M.~Guidal and P. Guichon, ``VGG code'', {\it private communication}. 

\bibitem{MuelScha} D.~Mueller and A.~Schafer,
  Nucl.\ Phys.\ B {\bf 739}, 1 (2006)

\bibitem{GuzPol} V.~Guzey and M.~V.~Polyakov,
  Eur.\ Phys.\ J.\ C {\bf 46}, 151 (2006)

\bibitem{Fox} G.~C.~Fox,
  Nucl.\ Phys.\ B {\bf 131}, 107 (1977).

\bibitem{Yndurain}F.~J.~Yndurain,
  Phys.\ Lett.\ B {\bf 74}, 68 (1978).
     
\bibitem{zan} M.~Gockeler {\it et al.} [QCDSF Collaboration],
  Phys.\ Rev.\ Lett.\  {\bf 92}, 042002 (2004); M.~Gockeler {\it et al.},
  Nucl.\ Phys.\ Proc.\ Suppl.\  {\bf 153}, 146 (2006)
  [arXiv:hep-lat/0512011]

\bibitem{LHPC} P.~Hagler, J.~Negele, D.~B.~Renner, W.~Schroers, T.~Lippert and K.~Schilling
                  [LHPC collaboration],
  Phys.\ Rev.\ D {\bf 68}, 034505 (2003).
 
\bibitem{part2} S.Ahmad, H.Honkanen, S.Liuti and S.K. Taneja, {\it ``Generalized Parton Distributions 
from Hadronic Observables: $\zeta \neq 0$}, in preparation.

%%%% SECTION II
\bibitem{MarGol} K.~J.~Golec-Biernat and A.~D.~Martin,
  Phys.\ Rev.\ D {\bf 59}, 014029 (1999)
 
\bibitem{BroDie} S.~J.~Brodsky, M.~Diehl and D.~S.~Hwang,
  Nucl.\ Phys.\ B {\bf 596}, 99 (2001)
  
\bibitem{MeyMul} H.~Meyer and P.~J.~Mulders,
  Nucl.\ Phys.\ A {\bf 528}, 589 (1991).
  
\bibitem{MelSchTho}W.~Melnitchouk, A.~W.~Schreiber and A.~W.~Thomas,
  Phys.\ Rev.\ D {\bf 49}, 1183 (1994)

\bibitem{LiuTan2} S.~Liuti and S.~K.~Taneja,
  Phys.\ Rev.\ C {\bf 72}, 034902 (2005)

\bibitem{BroCloGun} S.~J.~Brodsky, F.~E.~Close and J.~F.~Gunion,
  Phys.\ Rev.\ D {\bf 8}, 3678 (1973).
 
\bibitem{SzcLon} A.~P.~Szczepaniak and J.~T.~Londergan,
arXiv:hep-ph/0604266.
      
\bibitem{Bolz} J.~Bolz and P.~Kroll,
  Z.\ Phys.\ A {\bf 356}, 327 (1996)

\bibitem{DieKro2} M.~Diehl, T.~Feldmann, R.~Jakob and P.~Kroll,
  Nucl.\ Phys.\ B {\bf 596}, 33 (2001)
  [Erratum-ibid.\ B {\bf 605}, 647 (2001)]
  
\bibitem{HERMES} A.~Airapetian {\it et al.}  [HERMES Collaboration],
  Phys.\ Rev.\ Lett.\  {\bf 87}, 182001 (2001)
  
\bibitem{Jlab}S.~Stepanyan {\it et al.}  [CLAS Collaboration],
  Phys.\ Rev.\ Lett.\  {\bf 87}, 182002 (2001)
 
\bibitem{BroEst} S.~J.~Brodsky and F.~J.~Llanes-Estrada,
  Eur.\ Phys.\ J.\ C {\bf 46}, 751 (2006)

\bibitem{Schierholz} G.Schierholz, in {\it GPD 2006: Workshop on Generalized Parton Distributions}, June 2006,  
ECT$*$ Trento, Italy -- http://gpd.gla.ac.uk/gpd2006/index.php; and {\it private communication}.  

\bibitem{acha}   A.~Acha {\it et al.}  [HAPPEX collaboration],
 arXiv:nucl-ex/0609002.
 
\bibitem{JICR} C.~R.~Ji, Y.~Mishchenko and A.~Radyushkin,
  Phys.\ Rev.\ D {\bf 73}, 114013 (2006);
 H.~M.~Choi, C.~R.~Ji and L.~S.~Kisslinger,
  Phys.\ Rev.\ D {\bf 64}, 093006 (2001)
 
%%%%%%%%%% SECTION III
\bibitem{GSI} V. Barone, {\it et al.}, hep-ex 0505054.

\bibitem{Gluck} M. Gl\"{u}ck {\it et al.}, Z. Phys. {\bf C 41} (1989) 667; {\it ibid} 
Z. Phys. {\bf C 48} (1990) 471.
%\bibitem{GuzPol}   V.~Guzey and M.~V.~Polyakov,
%  Eur.\ Phys.\ J.\ C {\bf 46}, 151 (2006)

\bibitem{Camacho} C.~Munoz Camacho {\it et al.}  [Jefferson Lab Hall A Collaboration],
  arXiv:nucl-ex/0607029.  
%P. Bertin, in {\it GPD 2006: Workshop on Generalized Parton Distributions}, June 2006,  
%ECT$*$ Trento, Italy -- http://gpd.gla.ac.uk/gpd2006/index.php; and {\it private communication}. 

\bibitem{Pumplin}
  J.~Pumplin,
  AIP Conf.\ Proc.\  {\bf 792}, 50 (2005)

\bibitem{Jakob:1997wg}  
R.~Jakob, P.~J.~Mulders and J.~Rodrigues,
  Nucl.\ Phys.\ A {\bf 626}, 937 (1997)


   
%---------------- Data for G_Ep------------------------------------

%\cite{Price:1971zk}
\bibitem{GEP_exp}
  L.~E.~Price, J.~R.~Dunning, M.~Goitein, K.~Hanson, T.~Kirk and R.~Wilson,
%   ``Backward-angle electron-proton elastic scattering and proton
  %electromagnetic form-factors,''Z. Phys. {
  Phys.\ Rev.\ D {\bf 4} (1971) 45;
  %%CITATION = PHRVA,D4,45;%%;
%\cite{Simon:1980hu}
  G.~G.~Simon, C.~Schmitt, F.~Borkowski and V.~H.~Walther,
%   ``Absolute Electron Proton Cross-Sections At Low Momentum Transfer Measured
  %With A High Pressure Gas Target System,''
  Nucl.\ Phys.\ A {\bf 333}, 381 (1980).
  %%CITATION = NUPHA,A333,381;%%

%-------------- Data for G_Mp-----------------------------

%\cite{Hohler:1976ax}
\bibitem{GMP_exp}
  G.~Hohler, E.~Pietarinen, I.~Sabba Stefanescu, F.~Borkowski, G.~G.~Simon, V.~H.~Walther and R.~D.~Wendling,
  %``Analysis Of Electromagnetic Nucleon Form-Factors,''
  Nucl.\ Phys.\ B {\bf 114}, 505 (1976);
  %%CITATION = NUPHA,B114,505;%%
%\bibitem{Brash:2001qq}
  E.~J.~Brash, A.~Kozlov, S.~Li and G.~M.~Huber,
  %``New empirical fits to the proton electromagnetic form factors,''
  Phys.\ Rev.\ C {\bf 65}, 051001 (2002)
%  [arXiv:hep-ex/0111038].
  %%CITATION = HEP-EX 0111038;%%

%---------------- Data for G_En--------------------------------
%\cite{Eden:1994ji}
\bibitem{GEN_exp}
  T.~Eden {\it et al.},
%   ``Electric form-factor of the neutron from the H-2 (e (polarized), e-prime n
  %(polarized) ) H-1 reaction at Q**2 = 0.255-(GeV/c)**2,''
  Phys.\ Rev.\ C {\bf 50}, 1749 (1994);
  %%CITATION = PHRVA,C50,1749;%%
%\cite{Passchier:1999cj}
%\bibitem{Passchier:1999cj}
  I.~Passchier {\it et al.},
%   ``The charge form factor of the neutron from the reaction
  %H-2(pol.)(e(pol.),e' n)p,''
  Phys.\ Rev.\ Lett.\  {\bf 82}, 4988 (1999);
%  [arXiv:nucl-ex/9907012].
  %%CITATION = NUCL-EX 9907012;%%
%\cite{Herberg:1999ud}
%\bibitem{Herberg:1999ud}
  C.~Herberg {\it et al.},
%   ``Determination of the neutron electric form factor in the D(e,e' n)p
  %reaction and the influence of nuclear binding,''
  Eur.\ Phys.\ J.\ A {\bf 5}, 131 (1999);
  %%CITATION = EPHJA,A5,131;%%
%\cite{Golak:2000nt}
%\bibitem{Golak:2000nt}
  J.~Golak, G.~Ziemer, H.~Kamada, H.~Witala and W.~Gloeckle,
%   ``Extraction of electromagnetic neutron form factors through inclusive  and
  %exclusive polarized electron scattering on polarized He-3 target,''
  Phys.\ Rev.\ C {\bf 63}, 034006 (2001);
% [arXiv:nucl-th/0008008].
  %%CITATION = NUCL-TH 0008008;%%
%\cite{Madey:2003av}
%\bibitem{Madey:2003av}
  R.~Madey {\it et al.}  [E93-038 Collaboration],
%   ``Measurements of G(E)(n)/G(M)(n) from the H-2(e(pol.),e' n(pol.))H-1
  %reaction to Q**2 = 1.45-(GeV/c)**2,''
  Phys.\ Rev.\ Lett.\  {\bf 91}, 122002 (2003);
%  [arXiv:nucl-ex/0308007].
  %%CITATION = NUCL-EX 0308007;%%
%\cite{Warren:2003ma}
%\bibitem{Warren:2003ma}
  G.~Warren {\it et al.}  [Jefferson Lab E93-026 Collaboration],
%   ``Measurement of the electric form factor of the neutron at Q**2 =
  %0.5-GeV/c**2 and 1.0-GeV/c**2,''
  Phys.\ Rev.\ Lett.\  {\bf 92}, 042301 (2004);
%  [arXiv:nucl-ex/0308021].
  %%CITATION = NUCL-EX 0308021;%%
%\cite{Schiavilla:2001qe}
%\bibitem{Schiavilla:2001qe}
  R.~Schiavilla and I.~Sick,
  %``Neutron charge form factor at large q**2,''
  Phys.\ Rev.\ C {\bf 64}, 041002 (2001);
%  [arXiv:nucl-ex/0107004].
  %%CITATION = NUCL-EX 0107004;%%
%\cite{Rohe:1999sh}
%\bibitem{Rohe:1999sh}
  D.~Rohe {\it et al.},
%   ``Measurement of the neutron electric form factor G(en) at  0.67-(GeV/c)**2
  %via He-3(pol.)(e(pol.),e' n),''
  Phys.\ Rev.\ Lett.\  {\bf 83}, 4257 (1999).
  %%CITATION = PRLTA,83,4257;%%
%------------ Data for G_Mn----------------------------------
%\cite{Kubon:2001rj}
\bibitem{GMN_exp}
  G.~Kubon {\it et al.},
  %``Precise neutron magnetic form factors,''
  Phys.\ Lett.\ B {\bf 524}, 26 (2002);
%  [arXiv:nucl-ex/0107016].
  %%CITATION = NUCL-EX 0107016;%%
%\cite{Anklin:1994ae}
%\bibitem{Anklin:1994ae}
  H.~Anklin {\it et al.},
  %``Precision measurement of the neutron magnetic form-factor,''
  Phys.\ Lett.\ B {\bf 336}, 313 (1994);
  %%CITATION = PHLTA,B336,313;%%
%\cite{Anklin:1998ae}
%\bibitem{Anklin:1998ae}
  H.~Anklin {\it et al.},
  %``Precise measurements of the neutron magnetic form factor,''
  Phys.\ Lett.\ B {\bf 428}, 248 (1998);
  %%CITATION = PHLTA,B428,248;%%
%\cite{Lung:1992bu}
%\bibitem{Lung:1992bu}
  A.~Lung {\it et al.},
%   ``Measurements of the electric and magnetic form-factors of the neutron from
  %Q**2 = 1.75-GeV/c**2 to 4-GeV/c**2,''
  Phys.\ Rev.\ Lett.\  {\bf 70}, 718 (1993);
  %%CITATION = PRLTA,70,718;%%
%\cite{Xu:2000xw}
%\bibitem{Xu:2000xw}
  W.~Xu {\it et al.},
%   ``The transverse asymmetry A(T') from quasielastic polarized
  %He-3(pol.)(e(pol.),e') process and the neutron magnetic form factor,''
  Phys.\ Rev.\ Lett.\  {\bf 85}, 2900 (2000);
%  [arXiv:nucl-ex/0008003].
  %%CITATION = NUCL-EX 0008003;%%
%\cite{Xu:2002xc}
%\bibitem{Xu:2002xc}
  W.~Xu {\it et al.}  [Jefferson Lab E95-001 Collaboration],
%   ``PWIA extraction of the neutron magnetic form factor from quasi-elastic
  %He-3(pol.)(e(pol.),e') at Q**2 = 0.3-(GeV/c)**2 to 0.6-(GeV/c)**2,''
  Phys.\ Rev.\ C {\bf 67}, 012201 (2003);
%  [arXiv:nucl-ex/0208007].
  %%CITATION = NUCL-EX 0208007;%%
%\cite{Rock:1982gf}
%\bibitem{Rock:1982gf}
  S.~Rock {\it et al.},
%   ``Measurement Of Elastic Electron - Neutron Cross-Sections Up To Q**2 =
  %10-(Gev/C)**2,''
  Phys.\ Rev.\ Lett.\  {\bf 49}, 1139 (1982).
  %%CITATION = PRLTA,49,1139;%%
%------------- Data for G_Ep/G_Mp--------------------------------
%\cite{Jones:1999rz}
\bibitem{GEP_GMP}
  M.~K.~Jones {\it et al.}  [Jefferson Lab Hall A Collaboration],
%   ``G(E(p))/G(M(p)) ratio by polarization transfer in  e(pol.) p --> e
  %p(pol.),''
  Phys.\ Rev.\ Lett.\  {\bf 84}, 1398 (2000);
%  [arXiv:nucl-ex/9910005].
  %%CITATION = NUCL-EX 9910005;%%
%\cite{Gayou:2001qd}
%\bibitem{Gayou:2001qd}
  O.~Gayou {\it et al.}  [Jefferson Lab Hall A Collaboration],
  Phys.\ Rev.\ Lett.\  {\bf 88}, 092301 (2002);
%  [arXiv:nucl-ex/0111010].
  %%CITATION = NUCL-EX 0111010;%%
%\cite{Pospischil:2001pp}
%\bibitem{Pospischil:2001pp}
  T.~Pospischil {\it et al.}  [A1 Collaboration],
%   ``Measurement of G(E(p))/G(M(p)) via polarization transfer at Q**2 =
  %0.4-GeV/c**2,''
  Eur.\ Phys.\ J.\ A {\bf 12}, 125 (2001);
  %%CITATION = EPHJA,A12,125;%%
%\cite{Milbrath:1997de}
%\bibitem{Milbrath:1997de}
  B.~D.~Milbrath {\it et al.}  [Bates FPP collaboration],
%   ``A comparison of polarization observables in electron scattering from  the
  %proton and deuteron,''
  Phys.\ Rev.\ Lett.\  {\bf 80}, 452 (1998)
  [Erratum-ibid.\  {\bf 82}, 2221 (1999)]
%  [arXiv:nucl-ex/9712006].
  %%CITATION = NUCL-EX 9712006;%%

%------ End of data block -----------------

%\cite{Kelly:2004hm}
\bibitem{Kelly:2004hm}
  J.~J.~Kelly,
  %``Simple parametrization of nucleon form factors,''
  Phys.\ Rev.\ C {\bf 70}, 068202 (2004).
  %%CITATION = PHRVA,C70,068202;%%

\bibitem{Alekhin:2002fv}
  S.~Alekhin,  Phys.\ Rev.\ D {\bf 68}, 014002 (2003)

\bibitem{CTEQ} J.~Pumplin, D.~R.~Stump, J.~Huston, H.~L.~Lai, P.~Nadolsky and W.~K.~Tung,
  %``New generation of parton distributions with uncertainties from global  QCD
  %analysis,''
  JHEP {\bf 0207}, 012 (2002)

\bibitem{MRST} R.~S.~Thorne, A.~D.~Martin and W.~J.~Stirling,
  %``MRST parton distributions - status 2006,''
  arXiv:hep-ph/0606244.

\bibitem{SchThoLon}A.~W.~Schreiber, A.~W.~Thomas and J.~T.~Londergan,
  Phys.\ Rev.\ D {\bf 42}, 2226 (1990).

%%%%%%% SECTION IV
\bibitem{Soper} D.~E.~Soper,
  Phys.\ Rev.\ D {\bf 15}, 1141 (1977).

\bibitem{Burk2} M.~Burkardt,
 Phys.\ Rev.\ D {\bf 66}, 114005 (2002).

\bibitem{anselmino} M.~Anselmino, M.~Boglione, U.~D'Alesio, A.~Kotzinian, F.~Murgia and A.~Prokudin,
  Phys.\ Rev.\ D {\bf 72}, 094007 (2005)
  [Erratum-ibid.\ D {\bf 72}, 099903 (2005)]
  
\bibitem{Hagler}M.~Diehl and Ph.~Hagler,
   Eur.\ Phys.\ J.\ C {\bf 44}, 87 (2005).
  
\bibitem{Ava} H. Avakian, {\it private communication}.


\end{thebibliography}
\end{document}